\documentclass[a4paper,dvips,12pt]{article}
\usepackage{amsmath,amssymb,exscale}   
\usepackage{array,multicol}
\usepackage{afterpage,float,flafter}   
\usepackage{epsfig,rotating,pifont,fancybox}   
\usepackage{cite}
\usepackage{psfrag}
\usepackage{comment}
\makeatletter   
\@addtoreset{equation}{section}   
\makeatother

\newcommand\be{\begin{equation}}
\newcommand\ee{\end{equation}}
\newcommand{\brr}{\begin{eqnarray}}
\newcommand{\err}{\end{eqnarray}}
\newcommand{\Dslash}{\,/\!\!\!\!D}
\newcommand{\Id}{\text{\small 1}\hspace{-3.5pt}\text{1}}
\newcommand{\myfrac}[2]{\scriptstyle#1/#2}
\newcommand{\bea}{\begin{eqnarray}}   
\newcommand{\eea}{\end{eqnarray}}   
   
\newcommand{\NPB}[3] {Nucl.~Phys. \textbf{B#1} (#2) #3}   
\newcommand{\PLB}[3] {Phys.~Lett. \textbf{B#1} (#2) #3}

\title{   
\vspace*{-0.8cm}   
\begin{flushright}   
\normalsize{      
\texttt{hep-th/0305024}}\\ 
\end{flushright}    
\vspace{1cm}
\Large{\sc Localized anomalies in orbifold gauge 
theories~\footnote{Work supported in part by CICYT, Spain, under
contracts FPA 2001-1806 and FPA 2002-00748, and by EU under contracts
HPRN-CT-2000-00152 and HPRN-CT-2000-00148.}}
\vspace*{.5cm} \author{\large {\sc G.~v.~Gersdorff~$^\dagger$ and
M.~Quir{\'o}s}~$^{\ddagger\,\dagger}$\\ \\
\emph{$^\ddagger$~Instituci\'o Catalana de Recerca i Estudis
Avan\c{c}ats (ICREA)}\\ \emph{$^\dagger$~Depart.~of~Theoretical
Physics}\\ \emph{IFAE/UAB E-08193 Bellaterra, Spain}} } \date{}

\begin{document}
\maketitle
\thispagestyle{empty}
\vspace*{.5cm}

\begin{abstract}\noindent
We apply the path-integral formalism to compute the anomalies in
general orbifold gauge theories (including possible non-trivial
Scherk-Schwarz boundary conditions) where a gauge group $\mathcal G$
is broken down to subgroups $\mathcal H_f$ at the fixed points
$y=y_f$. Bulk and localized anomalies, proportional to
$\delta(y-y_f)$, do generically appear from matter propagating in the
bulk.  The anomaly zero-mode that survives in the four-dimensional
effective theory should be canceled by localized fermions (except
possibly for mixed $U(1)$ anomalies).  We examine in detail the
possibility of canceling localized anomalies by the Green-Schwarz
mechanism involving two- and four-forms in the bulk.  The four-form
can only cancel anomalies which do not survive in the 4D effective
theory: they are called globally vanishing anomalies. The two-form may
cancel a specific class of mixed $U(1)$ anomalies. Only if these
anomalies are present in the 4D theory this mechanism spontaneously
breaks the $U(1)$ symmetry.  The examples of five and six-dimensional
$\mathbb Z_N$ orbifolds are considered in great detail. In five
dimensions the Green-Schwarz four-form has no physical degrees of
freedom and is equivalent to canceling anomalies by a Chern-Simons
term.  In all other cases, the Green-Schwarz forms have some physical
degrees of freedom and leave some non-renormalizable interactions in
the low energy effective theory.  In general, localized anomaly
cancellation imposes strong constraints on model building.
\end{abstract}
\vspace{2.cm}   
   
\begin{flushleft}    
April 2003 \\   
\end{flushleft}
%%%%%%%%%%%%%%%%%%%%%%%%%%%%%%%%%%%%%%%%%%%%%%%%%%%%%%   
\newpage
\section{\sc Introduction}
\label{introduction}

The existence of extra dimensions (on top of the four space-time
coordinates) is a general feature of theories aiming to unify all
known interactions (including gravity) with quantum mechanics. In
particular string theories predict six, and M-theory seven, extra
dimensions and it is widely believed that all string theories are
related to each other by various string
dualities~\cite{polchinski}. Unlike in the case of the perturbative
heterotic string where all scales --the string ($M_s$) and the
compactification ($M_c$) scales-- are close to each other and to the
four-dimensional Planck scale, in some recent string constructions (as
e.~g. non-perturbative heterotic string or type I strings) it has been
shown that both the string length~\cite{lykken} and (some of the)
compactification radii can lie in the range of the inverse TeV length,
with possible interesting phenomenological
implications~\cite{antoniadis}. Moreover the existence of branes
(e.~g.~D-branes in type I strings~\cite{polchinski}) makes it possible
that the Standard Model fields propagate in a brane with $p$ {\it
longitudinal} dimensions spanning a $(4+p)$-dimensional ($p\geq 1$)
world-sheet while gravity propagates in the bulk of the
higher-dimensional space where the {\it transverse} dimensions can be
as large as the submillimeter size, thus ``explaining'' the smallness
of four-dimensional gravitational interactions~\cite{ADD}. In that
case, if there is a little hierarchy between the string and
compactification scales ($M_s/M_c\gg 1$) the Standard Model
interactions are described by an effective {\it field} theory in
$(4+p)$-dimensions with a cutoff at $\Lambda\simeq
M_s$~\cite{ABQ}. This opened up the exciting possibility of new
physics beyond the Standard Model scale and below the string scale
with large (TeV) extra dimensions and towers of Kaluza-Klein states
that can give rise to new phenomenology at future
colliders~\cite{pheno} and to new mechanisms for old phenomena as
supersymmetry and electroweak symmetry breaking~\cite{Tasi}. In
particular non-trivial compactification of the extra dimensions, as
orbifold compactification~\cite{orbifolds} and Scherk-Schwarz boundary
conditions~\cite{SS}, can shed some light on those problems.

The previous ideas gave rise to a plethora of field-theoretical models
in extra dimensions~\cite{alex,barbieri} where both phenomenological
and theoretical studies have recently been developed. However, unlike
in string constructions where the consistency of the theory is
guaranteed by the symmetries of the string (e.~g. modular invariance
in heterotic string models) in field-theoretical ones this consistency
must be imposed. In particular one of the main ingredients coming
automatically in string models, i~e. anomaly freedom, must be enforced
in field-theoretical models. This requisite is a very important one
since it is responsible for the consistency of the theory at the
quantum level. Moreover since anomalies are an infra-red phenomenon
their cancellation is a common requirement where both string and field
theory constructions should meet. Put differently, since anomalies are
generated by the massless spectrum they are a common feature of any
string theory and any effective field theory descending from it. To
conclude, anomaly cancellation in field theories in extra dimensions
is a requirement for the consistency of the corresponding theory at
the quantum level and also a requirement for it to descend from an
underlying string theory.

The question of anomaly cancellation in field-theoretical models in
the presence of extra dimensions has recently been addressed. The
first step in this direction was given in
Ref.~\cite{Arkani-Hamed:2001is} where a simple five-dimensional model
compactified on the orbifold $S^1/\mathbb Z_2$ was considered. It was
observed that the five-dimensional anomaly was localized at the
orbifold fixed points $y=0,\pi R$ while there was no bulk anomaly, a
simple consequence of the fact that the five-dimensional theory is
non-chiral. However cancellation of localized anomalies was
automatically induced by cancellation of anomalies in the
four-dimensional theory. This feature was proven not to be a general
one in subsequent studies of the orbifold $S^1/\mathbb
Z_2\otimes\mathbb
Z^\prime_2$~\cite{Scrucca:2001eb,Pilo:2002hu,Barbieri:2002ic,
GrootNibbelink:2002wv,GrootNibbelink:2002qp,Kim:2002ab} where
cancellation of anomalies in the four-dimensional theory did not imply
cancellation of localized anomalies. However it was proven that the
latter could be achieved by the introduction of a Chern-Simons
counterterm without affecting the four-dimensional effective
theory. The anomalies in particular higher-dimensional (string and
field-theoretical) models have been very recently under
consideration~\cite{Scrucca:2002is,Gmeiner:2002es,GrootNibbelink:2003gb,Asaka:2002,Lee:2002qn}.

The aim of this paper is to see how localized anomalies can be
computed in arbitrary orbifold models in any dimension and how the
cancellation by the Green-Schwarz mechanism generalizes to those
cases.  We consider orbifolds defined on an arbitrary
$(D-4)$-dimensional compact space $C$ modded out by an arbitrary
discrete group $\mathbb G$ acting non-freely on $C$ with fixed points
at $y=y_f$. Our purpose is to obtain general features of localized
anomalies and anomaly cancellation that could be held by any orbifold
construction. We assume an arbitrary gauge group in the bulk $\mathcal
G$ broken by the orbifold action to subgroups $\mathcal H_f$ at the
fixed points and with possible Scherk-Schwarz boundary conditions. We
have proved that, as expected, the absence of anomalies in the
four-dimensional theory of zero modes can not cope with the
cancellation of localized anomalies in arbitrary orbifolds.  We then
study the contribution to anomalies from Green-Schwarz (GS) $p$-forms
propagating in the bulk~\cite{Green:sg}.  It turns out that in general
two and four-forms (or their corresponding duals) can contribute to
localized anomalies and eventually cancel existing
ones~\footnote{Local anomaly cancellation with four-forms was first
discussed in Ref.~\cite{Scrucca:2002is}. Two-forms were recently
employed to cancel localized anomalies in heterotic string
constructions \cite{GrootNibbelink:2003gb}, a mechanism similar to the
one of Ref.~\cite{Horava:1996ma}.}. We will define {\it globally
vanishing} localized anomalies those whose zero-modes vanish upon
integration over the extra dimensions and thus do not survive in the
four-dimensional effective theory.  The localized anomalies whose
zero-modes survive in the four-dimensional theory are consequently
called globally non-vanishing anomalies.

The contribution from the two-form is always a mixed $U(1)$ anomaly,
which can be globally vanishing or not. In case it is globally
non-vanishing, the $U(1)$ will be spontaneously broken in the low
energy theory. In the four-dimensional (4D) limit we are left with
non-renormalizable interactions which depend on the
compactification. In 5D models we find an axion $a(x)$, coupling at
low energy like
\be
a\, \text{tr}\,F_{[\mu\nu}F_{\rho\sigma]}\epsilon^{\mu\nu\rho\sigma},
\ee
while in $6D$ compactifications we find in addition zero
mode-interactions of two-forms as
\be
C_{\mu\nu}\text{tr}\,F_{[\rho\sigma}F_{56]}\epsilon^{\mu\nu\rho\sigma}.
\ee

The contribution from the GS four-form gives all kind of pure gauge
anomalies, but no mixed $U(1)$ gravitational ones if $\mathcal G$ is
simple. Those anomalies are {\em always} globally vanishing, which is
consistent with the fact that non-abelian anomalies can not be
canceled by GS in 4D. As a consequence, GS cancellation of this kind
of anomalies can only work for anomalies which do not have any zero
mode, i.e.~are globally vanishing. In the 5D models, a four-form has
no physical degrees of freedom and can thus be integrated out
algebraically, leaving over precisely the CS counterterm found earlier
in the literature. In 6D, the four-form has an axionic degree of
freedom which in general compactifications will have a zero mode
$a(x)$, leaving the following non-renormalizable coupling at low
energy:
\be
a\, \text{tr}\,F_{[\mu\nu}F_{\rho\sigma}F_{56]}\epsilon^{\mu\nu\rho\sigma},
\ee
In both cases we will find that the localized anomalies generated by
the GS forms have a quite peculiar form, which considerably restricts 
the possibility of canceling the localized anomalies produced by bulk and
brane fermions.

The outline of this paper is as follows. In section~\ref{orbifold} the
different projectors that appear in the orbifold construction are
defined and the gauge structure localized at the different fixed
points described. Section~\ref{wilson} is devoted to the introduction
of Scherk-Schwarz non-trivial boundary conditions in the orbifold
structure. The equivalence between Scherk-Schwarz~\cite{SS} and
Hosotani~\cite{Hosotani:1983xw} breaking is not guaranteed (especially
in orbifolds in $D\geq 6$ dimensions) and the conditions for it are
explicitly established. An important case where this equivalence does
not hold is when the Scherk-Schwarz breaking proceeds through discrete
Wilson lines along the Cartan subalgebra, in which case the
corresponding orbifold can be described by a different one with
periodic fields defined on a larger torus modded out by a larger
discrete group. While the Hosotani mechanism spontaneously breaks the
gauge group in the zero mode sector but not the gauge groups localized
on the branes, the latter breaking is possible in the SS-mechanism and
naturally has an impact on the anomalies localized at the fixed
points.  In section~\ref{anomalies} the path-integral
method~\cite{Fujikawa:1979ay} is applied to compute the anomalies in
the previously defined orbifold structure. A bulk term, corresponding
to anomalies in the higher-dimensional gauge theory, and localized
terms on the orbifold fixed points are obtained. The structure of the
localized anomalies is disentangled in general. In section~\ref{GS}
the cancellation of localized anomalies by the Green-Schwarz mechanism
is studied. The Green-Schwarz $p$-forms in the bulk in general lead to
bosonic degrees of freedom in the 4D effective theory with
non-renormalizable interactions that could have some phenomenological
consequences. $\mathbb Z_N$ orbifolds in $D=5$ and $D=6$ are analyzed
in great detail in section~\ref{examples} whose results could be
easily applied to different orbifold and gauge constructions. Finally
section~\ref{conclusion} contains our conclusions and appendices A and
B some useful conventions and technical details about the orbifolds
studied in section~\ref{examples}.

\section{\sc Orbifold projectors}
\label{orbifold}

In this section we will consider a generic orbifold defined in $D-4$
extra dimensions and the corresponding projectors that will be used in
the calculation of the brane anomalies section~\ref{anomalies}.  The
starting point is the $C/\mathbb{G}$-orbifold $M^4\times
C/\mathbb{G}$, where $C$ is a compact $(D-4)$-dimensional manifold
(e.g.~a torus $T^{D-4}$) which is modded out by the discrete group
$\mathbb{G}$ that acts non-freely on $C$.  We will parametrize
$M^4\times C$ with the coordinates $x^M$ which we split into
$(x^\mu,x^i\equiv y^i)$, $\mu=0,\dots,3$ and $i=1,\dots,D-4$.  The
orbifold is constructed by identifying the orbits of $\mathbb{G}$,
i.e.
\be
x\sim P_k x
\label{orbspace}
\ee 
where the operators $P_k$ acting on $C$ are a representation of the
group elements $k\in \mathbb{G}$. Since $\mathbb{G}$ is acting
non-freely on $C$ it means that given $k\in \mathbb{G}$ there are some
points $y_{kf}\in C$ such that $P_k y_{kf}=y_{kf}$. The set
$\{y_{kf}\}$ are the fixed points of the $k$-sector of the orbifold.
For some orbifolds the set $\{y_f\}\equiv \cup_k\{y_{kf}\}$
constitutes a hypersurface inside $C$ in which case they are
correspondingly called fixed lines, planes, etc. We will call
generically these sets as fixed points.

The orbifold group $\mathbb{G}$ also has a representation on the
fields:
\be \phi(x)\rightarrow\lambda_k\otimes{\cal
P}_{k\,\sigma}~\phi(P_k^{-1}x)
\label{orbfields}
\ee
where ${\cal P}_{k\, \sigma}$ acts on the fermionic indices of the
field $\phi$, determined by the spin $\sigma$, and $\lambda_k$ acts on
internal gauge and flavor indices. We assume that the theory in the
bulk of the extra dimensions is a D-dimensional gauge theory with
gauge group $\mathcal G$ and the field $\phi$ transforms as an
(irreducible) representation $R$ of $\mathcal G$. The matrices ${\cal
P}_{k\,\sigma}$ and $\lambda_k$ must form representations of
$\mathbb{G}$.  According to the identification of Eq.~(\ref{orbspace})
we then have~\footnote{For the moment we are only considering fields
$\phi$ with trivial (periodic) boundary conditions on $C$. Twisted
(Scherk-Schwarz) boundary conditions will be introduced in
section~\ref{wilson}.}
\be
\phi(x)=\lambda_k\otimes{\cal P}_{k\,\sigma}~\phi(P_k^{-1}x),
\qquad\forall k\in {\mathbb G}
\label{orbfields2}
\ee
These are the $N\equiv|G|$ orbifold constraints the fields have to satisfy.

We now introduce the action of $\mathbb{G}$ on function space by
defining the unitary operator
\be
\hat P_k |x\rangle=|P_kx\rangle,\qquad
\langle x|\hat P_k=\langle P_k^{-1}x|.
\ee
Eq.~(\ref{orbfields2}) can then be rewritten as a linear operation on
the space of fields on the torus, $|\phi\rangle$ (where $\phi(x)\equiv
\langle x|\phi\rangle$), as
\be
(1-\lambda_k\otimes{\cal P}_{k\,\sigma}\otimes\hat P_k)|\phi\rangle=0,
\qquad\forall k\in {\mathbb G}.
\label{orbfields3}
\ee

We are looking for a projector $Q_\phi$ which, acting on a generic
field $\phi$ on $C$ gives a field which satisfies
Eq.~(\ref{orbfields3}). In other words $Q_\phi$ should satisfy the
conditions
\be
(1-\lambda_k\otimes{\cal P}_{k\,\sigma}\otimes\hat P_k)Q_\phi=0,
\qquad\forall k\in {\mathbb G}.
\ee
One can easily check that $Q_\phi$ is given by
\be
Q_\phi=
\frac{1}{N}\sum_k\lambda_k\otimes{\cal P}_{k\,\sigma}\otimes\hat P_k,
\ee
where the factor $1/N$ guarantees that $Q^2_\phi=Q_\phi$ and we also
have $Q_\phi^\dagger=Q_\phi$.  Note that the sum includes also the
unit element.  A further property of $Q_\phi$ is that it is gauge
invariant $g^{-1}Q_\phi g=Q_\phi$ since it commutes with $\xi(x)\equiv
\xi^A(x)T^A$, $g=\exp\{i\xi(x)\}$.  For later use we also note that
given a projection $Q_\psi$ for a fermion $\psi$, such that
$Q_\psi\psi$ defines a constrained spinor, the projector for the Dirac
conjugate $Q_{\bar\psi}$ is related to it as
\be
Q_{\bar\psi}=\Gamma^0Q_\psi\Gamma^0,
\label{Qpsibar}
\ee
and $\bar\psi Q_{\bar\psi}$ defines the corresponding constrained
spinor on the orbifold.

In general, the orbifold action of $k\in \mathbb{G}$ can break the
gauge group $\mathcal G=\{T^A\}$ in the bulk to a subgroup
$\mathcal{H}_f$ at the fixed point $y_f$. We assign a ${\mathbb G}$
transformation to the generators according to
\be
T^A\rightarrow\Lambda_k^{AB}T^B,
\label{automorphism}
\ee
which must be an automorphism on the algebra, i.e.~leave the structure
constants invariant.  Whenever we can write this transformation as a
group conjugation, i.e.~$\Lambda_k^{AB}T^B=g_kT^Ag_k^{-1}$, this is
called an inner automorphism. It takes a simple structure in terms of
the Weyl-Cartan basis $\{T^A\}=\{E_\alpha,H_j\}$ where we can always
choose $g_k$ to be of the form $g_k=\exp(-2\pi i \vec V_k\cdot \vec
H)$ where $\vec V_k$ is the rank($\mathcal G$)-dimensional
twist-vector that defines the orbifold breaking. We then simply have
$\Lambda^{ij}=\delta^{ij}$, $\Lambda_k^{i\alpha}=0$ and
$\Lambda^{\alpha\beta}=\exp\{ -2\pi i\, \vec{\alpha}\cdot
\vec{V}_k\}\,\delta^{\alpha\beta}$ where $\vec{\alpha}$ is the
rank($\mathcal G$)-dimensional root associated to the generator
$E_\alpha$.  If Eq.~(\ref{automorphism}) cannot be written as a group
conjugation it is called an outer automorphism. An important example
is the ${\mathbb Z}_2$ outer automorphism which takes $T_A\rightarrow
-T_A^T$.

After choosing the breaking pattern $\Lambda_k$, the gauge bosons
satisfy Eq.~(\ref{orbfields2}) with $\lambda_k=\Lambda_k$ and
${\mathcal P}_{k,\,1}=P_k$.  At each fixed point $y_f$, the elements
$k\in \mathbb{G}$ which leave $y_f$ fixed form a subgroup
$\mathbb{G}_f\subset\mathbb{G}$.  By diagonalizing $\Lambda$, the
unbroken gauge group at each fixed point is $\mathcal H_f=\{T^A~|~
\Lambda_k^{AA}=1\ \forall k\in \mathbb{G}_f\}$ since only gauge bosons
$A_\mu^A$ corresponding to these generators are non-vanishing at
$y_f$. Furthermore, the unbroken gauge group in the effective
four-dimensional theory is given by $\mathcal H=\{T^A~|~
\Lambda_k^{AA}=1\ \forall k\in \mathbb{G}\}$ since this defines the
set of massless 4D gauge bosons.  We will simply write $\mathcal
H=\{T^a\}$ and $\mathcal{G/H}=\{T^{\hat a}\}$.  Clearly, ${\mathcal
H}=\cap_f{\mathcal H}_f$, so in particular ${\mathcal H}$ is a
subgroup of all ${\mathcal H}_f$. We will denote the generators of
$\mathcal{H}_f=\{T^{a_f} \}$ and
$\mathcal{G}/\mathcal{H}_f=\{T^{\hat{a}_f} \}$, where the set
$\{T^{\hat{a}_f} \}$ is a subset of $\{T^{\hat a}\}$. In particular
for fixed points such that $\mathbb{G}_f=\mathbb{G}$, one gets
$\mathcal{H}_f=\mathcal{H}$.

If there are matter fields transforming in some representation of
$\mathcal G$, we must have
\be
\lambda_kT^A\lambda^{-1}_k=\Lambda_k^{AB}T^B.
\label{condlambda}
\ee
in order to get invariance of the action under $\mathbb G$.  Note that
if the automorphism Eq.~(\ref{automorphism}) is an inner one, this
condition is satisfied by just identifying $\lambda_k$ with $g_k$
evaluated in the appropriate representation. If the automorphism is an
outer one, there might be restrictions on the representations in order
to find such a $\lambda_k$.  Note that the projectors $\lambda_k$ have
to satisfy the commutation property with the unbroken generators
$T^{a_f}$,
\be
[\lambda_k, T^{a_f}]=0.
\label{condlambda2}
\ee

\section{\sc Scherk-Schwarz versus Hosotani breaking}
\label{wilson}
In this section we will consider the case of general orbifolds in the
presence of Scherk-Schwarz~\cite{SS} boundary conditions and their
relation to the Hosotani~\cite{Hosotani:1983xw} mechanism. We will
analyze the conditions under which those two breakings are equivalent
and find the cases where they are not, with the subsequent impact on
the possible localized anomalies in particular models.

In theories with non-simply connected internal dimensions, as
orbifolds, fields may possess non-trivial boundary conditions when
moving along a closed but non-contractible cycle. Invariance of the
action is guaranteed as long as the resulting multiple values of the
field are related by an internal (local or global) symmetry
transformation (twist):
\be
\phi(y^i+2\pi n^i)=W(n^i)\phi(y^i),\qquad W(n^i)=\exp(2\pi i n^i\Omega_i).
\label{SSbc}
\ee
The boundary conditions defined in Eq.~(\ref{SSbc}) are known as
Scherk-Schwarz (SS) boundary conditions~\cite{SS}.  Here $\vec{n}$ is
a lattice vector, $W(\vec{n})$ is the Wilson line along the direction
$\vec{n}$ and $\Omega_i=\omega_i^AT^A$ where $T^A$ are the generators
of the internal symmetry and $\omega_i^A$ the SS-parameters. To make
sense out of the boundary condition~(\ref{SSbc}) one has to demand
that Wilson lines along different directions
commute~\footnote{Usually, consistency of the orbifold-- and
SS--boundary conditions puts further constraints on $\Omega_i$.},
i.~e.
\be\label{consW}
[W(n^i),W(m^i)]=0\,\Longleftrightarrow\,[\Omega_i,\Omega_j]=0\ .
\ee
The corresponding symmetry is then broken in the presence of the
Wilson lines $W(\vec{n})$. Notice that condition (\ref{consW}) is
trivially satisfied for the case of one extra dimension ($D=5$). For
higher-dimensional theories ($D\geq 6$) it imposes a non-trivial
restriction on the SS-breaking patterns.

In the case of a local symmetry, one can sometimes undo this twist by
means of a non-periodic gauge transformation which only depends on the
extra coordinates, $g(y)$. All fields become then periodic but some
extra components of the gauge fields acquire a vacuum expectation
value (VEV).  This is known as the
Hosotani-mechanism~\cite{Hosotani:1983xw}. The obvious guess to
achieve periodic fields is the gauge transformation
$g(y)=\exp(-iy^i\Omega_i)$ and one immediately obtains $\langle
A_i\rangle=\Omega_i$. However, a constant VEV for $A_i$ is only
possible if this configuration is left invariant by the orbifold
action, thus obtaining the constraint~\footnote{In this section we
will be considering for definiteness the case where the orbifold
breaking of the gauge group is by an inner automorphism, although some
of its conclusions could also be generalized to arbitrary
automorphisms.}
\be
\langle
A_i\rangle=P_i^j\lambda_k \langle A_j\rangle \lambda_k^{-1}.
\label{zeromodecond}
\ee
where $P_i^j$ is a matrix representation acting over the space indices
of the orbifold element $P_k$.  Obviously only SS--breaking satisfying
the parallel constraint
\be
\Omega_i=P_i^j\lambda_k \Omega_j \lambda_k^{-1}.
\label{omegacond}
\ee
can be given a Hosotani interpretation.  Note that the latter one is
the same constraint $\Omega_i$ has to satisfy if the gauge
transformation is to be consistent with the orbifold,
i.~e.~\footnote{This equation is the finite version of the statement
that the gauge parameters $\xi_A(y)$ must have the correct
transformation under the action of the orbifold group.} if condition
\be
g(P_k y)=\lambda_kg(y)\lambda_k^{-1}.
\label{gaugeconsistent}
\ee
holds. This equation is precisely fulfilled by
$g(y)=\exp(-iy^i\Omega_i)$ if $\Omega_i$ satisfies
Eq.~(\ref{omegacond}). For the case of one extra dimension the
$\mathbb Z_2$ orbifold action is $P^5_5=-1$ and
condition~(\ref{omegacond}) on $\Omega\equiv\Omega_5$ gives
$\{\Omega,\lambda\}=0$, or equivalently $W\lambda W=\lambda$ which
provides the usual consistensy condition on possible twist operators
$\lambda$ in the $S^1/\mathbb Z_2$ orbifold~\cite{alex}.

On the other hand, starting from the Hosotani mechanism in the
presence of a constant background $\langle A_i\rangle$ sometimes one
can also transform periodic fields into fields satisfying the
Scherk--Schwarz boundary conditions. In fact a constant VEV can only
be gauged away provided that 
\be\label{efe0}
F_{ij}=[A_i,A_j]=0
\ee
in agreement with the condition (\ref{consW}). Notice that again [as
it happened with the consistency of the SS-boundary conditions in
(\ref{consW})] for the case of one extra dimension the condition
(\ref{efe0}) is trivially satisfied, which shows that for the $D=5$
case the Hosotani breaking can always be intepreted as a
Scherk-Schwarz~\cite{SSH1} breaking~\footnote{The converse is not
true, i.~e. a Scherk-Schwarz breaking in a five-dimensional theory is
not necessarily interpreted as a Hosotani breaking if $A_5=\Omega_5$
does not have a zero mode.}. Again for higher-dimensional theories
($D\geq 6$) the condition (\ref{efe0}) is non-trivial.  We summarize
the relation between the SS and Hosotani mechanisms in
Fig.~\ref{hosotaniss}.
\begin{figure}[htb]
\be
%\fbox{\parbox{15.5cm}
\left\{
\begin{array}{c}
\fbox{\text{Hosotani}}\\ \\
\phi(y^i+2\pi n^i)=\phi(y^i)\\
\langle A_i\rangle=const\\
\langle
A_i\rangle=P_i^j\lambda_k \langle A_j\rangle \lambda_k^{-1}
\end{array}
\right\}
\begin{array}{c}
\xrightarrow{\hspace{.4cm}{[A_i,A_j]=0}\hspace{.4cm}}\\
\xleftarrow[\Omega_i=P_i^j\lambda_k \Omega_j \lambda_k^{-1}]{}
\end{array}
\left\{
\begin{array}{c}
\fbox{\text{Scherk--Schwarz (Wilson--Line)}}\\ \\
\phi(y^i+2\pi n^i)=\exp(2\pi i n^i\Omega_i)\phi(y^i)\\
\langle A_i\rangle=0\\
\left[\Omega_i,\Omega_j\right]=0
\end{array}
\right\}
%}
\ee
\caption{\it The relation between SS and Hosotani breaking. The two
schemes are equivalent if the respective constraints specified in the
lowest row can be satisfied simultaneously. In this case there exists
a non-periodic gauge transformation which gives $\Omega_i=\langle
A_i\rangle$.}
\label{hosotaniss}
\end{figure}

The Scherk-Schwarz boundary conditions change the periodicity of the
fields and thus in principle the surviving gauge groups ${\mathcal
H}_f$ in the branes, which in turn determines the possible localized
matter content. To determine the unbroken gauge group at a given fixed
point $y_f$ in the presence of SS-twists, one has to identify the
gauge fields whose boundary conditions do not force them to vanish at
$y_f$. Since fields are in general non-periodic, we have to work on the
covering space of the torus where the boundary conditions applied at
the fixed point $y_f$ give
\be
A_\mu(P_ky_f)\equiv A_\mu(y_f+2\pi n_{f,k})=
W(n_{f,k})A_\mu(y_f)W^{-1}(n_{f,k})=\lambda_kA_\mu(y_f)\lambda_k^{-1}
\ee
Here $k$ is restricted to the orbifold subgroup ${\mathbb G}_f$
defined at the end of the previous section. The second equality
reflects the SS boundary condition, while the last one the orbifold
boundary condition.  The unbroken subgroup ${\mathcal H}_f^W$ is thus
spanned by the generators that commute with
$W^{-1}(n_{f,k})\lambda_k$. Since, in the absence of SS--boundary
conditions, the unbroken subgroup at the fixed point $y_f$ ($\mathcal
H_f$) was defined as the subgroup commuting with $\lambda_k$, $k\in
\mathbb G_f$, it is clear that generically ${\mathcal H}_f^W$ will be
different from $\mathcal H_f$. However if the SS-breaking can be given
a Hosotani interpretation we will see that they coincide,
i.~e. ${\mathcal H}_f^W=\mathcal H_f$.  In fact using
Eq.~(\ref{gaugeconsistent}) one can easily check that
$W^{-1}(n_{f,k})\lambda_k=g^{-1}(y_f)\lambda_k g(y_f)$, where
$g(y)=\exp(-i y^i\Omega_i)$ is the gauge transformation which relates
the SS and Hosotani pictures. Consequently, when switching on the SS
twist, the brane gauge group ${\mathcal H}_f^W$ becomes
$g^{-1}(y_f){\mathcal H}_fg(y_f)$ and thus is equivalent to the one
without SS-breaking. At the end this is expected, since in the
Hosotani picture all fields remain periodic and so ${\mathcal H}_f$ is
unaffected by the VEV $\langle A_i\rangle$.

However, the brane gauge group can change when there is a
Scherk-Schwarz breaking that can not be given a Hosotani
interpretation. This case is most commonly referred to as ``discrete
Wilson line'' breaking. A trivial way to satisfy the consistency condition
(\ref{consW}) it to choose the Wilson line generators to lie in the
Cartan subalgebra, 
\be\label{discreteW}
\Omega_i=\omega_i^IH_I.
\ee
Let us look at the example of a torus modded out by the cyclic group
$\mathbb{G=Z}_N$. The action on the extra coordinates $y$ is given by
$P^n$, $0\leq n<N$ with $P^N=1$, where $P_i^j$ does not contain unit
eigenvalues. Then in order to have a Hosotani interpretation
[Eq.~(\ref{omegacond})] one would have to satisfy the equation $P_i^j
\Omega_j=\Omega_i$, a condition that can not be fulfilled since
$\det(P-1)\neq 0$~\footnote{ Another way of understanding this result
is the following. If the orbifold breaking of the gauge group is by an
inner automorphism, then necessarily the gauge fields $A_\mu^I$ have
zero modes, while $A_i^I$ do not and can not consequently acquire a
constant VEV.}.  Using the fact that $\sum_n P^n=0$, the purely
geometrical identity
\be
(\tau(n^i)P)^N=1,\qquad \tau(n^i):y^i\rightarrow y^i+2\pi n^i.
\ee
is satisfied.  This identity is reflected on the fields and one thus
obtains $(W(n^i)\lambda)^N=W^N\lambda^N=W^N=1$ which implies that
$N\omega_i^I=\text{integer}$ and the Wilson lines turn out to be
discrete.  As a concrete example consider $S^1/{\mathbb Z}_2$ with
${\mathcal G}=SU(3)$ broken down to ${\cal H}={\cal H}_0={\cal
H}_\pi=SU(2)\otimes U(1)$ by the inner automorphism characterized by
the twist vector $V=({\frac{1}{2}},0^7)$ or equivalently
$\lambda=\text{diag}(-1,-1,1)$. Taking $\omega_5^I=V^I$ the new gauge
group ${\cal H}_\pi$ is $SU(3)$, since $W^{-1}(-1)\lambda=1$ now
commutes with all generators.

In most cases it will be possible to find an equivalent description in
terms of a different orbifold without Wilson lines.  In particular,
the relation $W^N=1$ implies that on the bigger torus $T^{\prime p}$
with radius $R'=NR$ there are no Wilson lines. One obtains the same
physical space (and the same physical theory) by modding out
$T^{\prime p}$ with the bigger orbifold group $\mathbb G'$ generated
by $\{P,\tau(\hat n_i)\}$, where $\hat n_i$ are the basis vectors
defining the lattice of $T^p$. In the above example this amounts to
the orbifold group $\mathbb Z_2\otimes\mathbb Z_2'$
\cite{barbieri} defined on the circle with twice the radius.

\section{\sc Anomalies in orbifolds}
\label{anomalies}

In this section we will apply the path-integral method of anomaly
evaluation~\cite{Fujikawa:1979ay} to compute the anomalies in orbifold
gauge theories.  In chiral gauge theories anomalies arise if the
measure in the functional integral is not invariant under gauge
symmetry transformations $\xi(x)$:
\be 
{\cal D}\psi{\cal D}\bar\psi\rightarrow {\cal D}\psi{\cal
D}\bar\psi\exp(i{\cal A}).  
\label{anomaly}
\ee
where $\psi$ is a fermion propagating in the bulk of the D-dimensional
theory.  Furthermore the Lagrangian $\mathcal{L}_D$ changes as
\be
\mathcal{L}_D\to \mathcal{L}_D+\xi\cdot \mathcal{D}_M\cdot J^{M}(x)
\ee
where $\mathcal{D}$ is the covariant derivative in the adjoint
representation of the gauge group and $J_M^A(x)$ is the fermionic
current. Non-conservation of the current $J_M^A$ is provided by the
anomaly $\mathcal A$ in (\ref{anomaly}) by imposing $\xi$-invariance
of the generating functional of Green functions, as
\be \left(\mathcal{D}_M J^{M}\right)^A(x)=\frac{\delta\,\mathcal A}{\delta
\xi^A(x)}
\ee

The gauge transformation of the fermion fields in the
functional integral is defined as
\begin{align}
\psi(x)\rightarrow& (1+i\xi(x))\psi(x)\nonumber\\
\bar\psi(x)\rightarrow&\bar\psi(x)(1-i\xi(x))
\label{transf}
\end{align}
and the measure transforms with the (inverse) Jacobian, 
\begin{align}
{\cal D}\psi\rightarrow&
%\frac{1}{\det(1+i\xi)}{\cal D}\psi=
\exp(-i\,\text{tr}\,\xi){\cal D}\psi\nonumber\\
{\cal D}\bar\psi\rightarrow&
%\frac{1}{\det(1-i\xi)}{\cal D}\bar\psi= 
\exp(+i\,\text{tr}\,\xi){\cal D}\bar\psi\ .
\label{t4ransf2}
\end{align}
The trace goes over all degrees of freedom of $\psi$ including an
integration over spacetime.  If the fermions have chirality and/or
orbifolds constraints, one has to introduce appropriate projectors to
take the trace over the unconstrained spinors leading, for even
dimensions where chiral spinors are eigenstates of the Dirac matrix
$\Gamma^{D+1}$, to
\begin{align}
{\cal A}&=-\text{tr}\,\left\{
\xi\, Q_\psi \frac{1\pm\Gamma^{D+1}}{2}\right\}
+\text{tr}\,\left\{
\xi\, Q_{\bar\psi} \frac{1\mp\Gamma^{D+1}}{2}\right\}
\label{Deven}
\end{align}
For odd dimensions there is no notion of chirality, so we get
\begin{align}
{\cal A}&=-\text{tr}\,
\xi\, Q_\psi +\text{tr}\,
\xi\, Q_{\bar\psi}
\label{Dodd}
\end{align}
These traces are to be regularized with e.g.~an exponential, so we
get the final result for even and odd D respectively
\begin{align}
{\cal A}&=-\lim_{M\rightarrow\infty}\text{tr}\left\{ \xi\left( Q_\psi
\frac{1\pm\Gamma^{D+1}}{2}-
\Gamma^0Q_\psi\Gamma^0\frac{1\mp\Gamma^{D+1}}{2}\right)
\exp(-\Dslash^2/M^2)\right\}\\
{\cal A}&=-\lim_{M\rightarrow\infty}\text{tr}\left\{ \xi\left( Q_\psi
-\Gamma^0Q_\psi\Gamma^0\right)\exp(-\Dslash^2/M^2)\right\}
\end{align}
where we have made use of the relation in Eq.~(\ref{Qpsibar}).

Evaluating these traces is fairly straightforward.  Using the identity
\be
\Dslash^2=D^2-\frac{ig}{2}F_{MN}\Gamma^{MN}
\ee
one can expand $e^{-\Dslash^2/M^2}$ in powers of $\Gamma^{MN}$.  For
even D-dimensional chiral fermions ($\sigma=1/2$) the traces over
gauge and Dirac indices and function space factorize as in the case of
smooth manifolds, but now involve insertions coming from the presence
of the projector $Q_\psi$. Then~\footnote{For odd $D$ or for
Dirac-fermions in even $D$ one should just remove the
chirality-projectors.}
\begin{align}
&\mathcal{A}=-\lim_{M\to\infty}\sum_{k,r}\frac{1}{r!}\left(
\frac{ig}{2M^2} \right)^r \text{tr}
\left[ \left( {\cal
P}_{k\frac{1}{2}} \frac{1\pm\Gamma^{D+1}}{2}- \Gamma^0 {\cal
P}_{k\frac{1}{2}}\Gamma^0\frac{1\mp\Gamma^{D+1}}{2}\right)
\Gamma^{M_1N_1}\dots\Gamma^{M_rN_r}\right]\nonumber\\ &\int
d^Dx~
\left[ \xi^A(x) \langle x|\hat P_k \exp(-\partial^2/M^2)
|x\rangle\, F^{B_1}_{M_1N_1}(x)\dots F^{B_r}_{M_rN_r}(x)
\right]
\text{tr}\left[ T^A\lambda_k T^{B_1}\dots
T^{B_r} \right]
\label{anomalia}
\end{align}

In each term in the sum over $r$ we have already neglegted terms which
are suppressed by inverse powers of $M$ greater than $2r$ that come
from the exact expansion of $\exp(\Dslash^2/M^2)$.  The insertion of
$\hat P_k$ in the integration over $x$ requires a brief examination.
In case $k=id$ the identity in $\mathbb G$ --recall that $Q$ always
contains such a term-- we are simply computing the conventional
$D$-dimensional bulk anomaly and we will have to evaluate the matrix
element in the usual way. We only give the result in even $D$:
\be
\langle x|\exp(-\partial^2/M^2)|x\rangle=
i\frac{M^D}{2^{D}\pi^{D/2}}
\label{matrixelement1}
\ee
This result plugged into Eq.~(\ref{anomalia}) selects in the
$M\to\infty$ limit the term corresponding to $r=D/2$ that provides the
bulk anomaly. In fact it is easy to check that only the term $r=D/2$
survives the $M\to\infty$ limit. The terms with $r>D/2$ go to zero as
$M^{D/2-r}$ while those with $r<D/2$ vanish by the properties of
$2^{D/2}$-dimensional Dirac matrices. For odd $D$ the identity in
$\mathbb G$ provides a vanishing contribution to the anomaly because
the prefactor in (\ref{anomalia}) is zero, as it should be since the
theory in the bulk is not chiral.  For the case $k\neq\,id$ we obtain
brane anomalies due to the insertion of a non-trivial $\hat P_k$. The
matrix element can be split into a four-dimensional factor times a
$(D-4)$-dimensional one~\footnote{We assume here that the subspace
left invariant by $P_k$ is four-dimensional. The generalization to
higher-dimensional fixed points is a trivial task.}
\begin{align}
\langle x|\hat P_k \exp(-\partial^2/M^2) |x\rangle&=
\langle x^\mu|\exp(-\partial_\mu\partial^\mu/M^2)|x^\mu\rangle
\langle y|\hat P_k \exp(-\partial_i\partial^i/M^2)|y\rangle
\label{matrixelement}
\end{align}
The first factor can be read off from Eq.~(\ref{matrixelement1}) for
$D=4$. The second factor is finite in the limit $M\rightarrow\infty$
and is computed to be~\footnote{To alleviate the
notation we are using $\delta(y)=\Pi_{i=1}^{D-4}\delta(y^i)$.}
\begin{align}
\langle y|\hat P_k
|y\rangle=\delta(y-P_ky)=
\frac{1}{|\det(1-P_k)|}\sum_f\delta(y-y_f).
\end{align}
where the sum runs over all the fixed points of $P_k$. The
determinant~\footnote{Here it is understood that $P_k$ is restricted to
the extra-dimensional space to render the determinant non-zero.} is
equal to the number $\nu_k$ of those fixed points according to
Lefschetz' theorem 
\be
\nu_k=\left|\det(1-P_k)\right|
\label{lefschetz}
\ee
This identity can be easily shown by taking the trace of $\hat P_k$ in the
subspace spanned by $|y\rangle$. Evaluated in position space this
gives $\nu_k/|\det(1-P_k)|$ while evaluation in momentum space gives
\be
\sum_{\vec \ell}\langle \vec \ell|\hat P_k
|\vec \ell\rangle=\sum_{\vec \ell} \delta_{\vec \ell,P\vec \ell}=1.
\ee
The final result for the matrix element is (for $k\neq
\text{id}$),
\be 
\langle x|\hat P_k \exp(-\partial^2/M^2) |x\rangle=
i\frac{M^4}{16\pi^2}\frac{1}{\nu_k}\sum_f\delta(y-y_f)+{\cal
O}(M^2).  
\label{otra}
\ee

Replacing (\ref{otra}) into (\ref{anomalia}) selects, in the limit
$M\to\infty$, the term corresponding to $r=2$ that gives rise to
localized anomalies at the orbifold fixed points $y_f$.  Since
$\lambda_k$ in (\ref{anomalia}) commute with all $T^{a_f}$ for $k\in
{\mathbb G}_f$, they are just constants on each irreducible
representation space of $\mathcal{H}_f$ and can thus be taken out of
the trace.  In fact the localized anomaly at the fixed point $y_f$
coming from a fermion in the representation $\mathcal{R_G}$ of
$\mathcal G$ with branching rule $\mathcal{R_G}=\oplus_i \mathcal
R^i_{\mathcal H_f}$ can be written as
\be\label{extra}
\mathcal A_f=-\frac{g^2}{32 \pi^2}\int d^Dx\, \xi^{a_f}\,
F^{b_f}_{\mu\nu}\,F^{c_f}_{\rho\sigma}\,
\epsilon^{\mu\nu\rho\sigma}\sum_ic_{if}\ {\rm str}_i
\left(T^{a_f} T^{b_f} T^{c_f}\right) \delta(y-y_f) 
\ee
where str{$_i$} denotes the symmetrized trace in the representation $\mathcal
R^i_{\mathcal H_f}$ and the $c_{if}$ are orbifold coefficients coming
from the evaluation of Eq.~(\ref{anomalia}).  Explicit calculation of
these coefficients in five and six-dimensional orbifolds will be
provided in section~\ref{examples}. If there are localized fermions
$\psi_f$ at the orbifold fixed point $y_f$ the usual four-dimensional
methods~\cite{Fujikawa:1979ay} yield the additional localized
anomalies $\mathcal{A}_f$ given by
\be \mathcal{A}_f=-\frac{g^2}{32\pi^2}\int d^Dx\, \xi^{a_f}\,
F^{b_f}_{\mu\nu}\,F^{c_f}_{\rho\sigma}\,\epsilon^{\mu\nu\rho\sigma}\ {\rm str}
\left(T^{a_f} T^{b_f} T^{c_f}\right) \delta(y-y_f) 
\label{anomlocal}
\ee

In the case there are $U(1)$ gauge bosons $A_\mu^{\alpha_f}$ in
$\mathcal H_f$ at $y=y_f$ there can also be localized mixed $U(1)$
gravitational anomalies from the non-invariance of the fermionic
determinant in the presence of a background gravitational field, that
can be obtained using functional methods as we did in
section~\ref{anomalies}. We expect the orbifold projection to generate
at the orbifold fixed point $y_f$ the localized anomalies,
\begin{equation}\label{u1grav}
\mathcal A_f^{U(1)-grav}=-\frac{1}{384\pi^2}\int
d^Dx\,\xi^{\alpha_f}\,\left(\sum_i c_{if}d_iq_i\right)\,
\frac{1}{2}\epsilon^{\mu\nu\rho\sigma} R^{\alpha\beta}_{\
\ \mu\nu}R_{\alpha\beta\rho\sigma}\ \delta(y-y_f)
\end{equation}
where $d_i$ is the dimension of the representation with charge $q_i$,
$R_{\mu\nu\rho\sigma}$ is the 4D Riemann-Christoffel tensor induced by
the higher-dimensional gravitational background, and $c_{if}$ are the
orbifold coefficients. In the particular case where all coefficients
$c_{if}$ for the different fields at a given fixed point are equal we
can take $c_{if}$ out of the sum and the condition for local anomaly
cancellation becomes the familiar one ${\rm tr}\,Q=0$.

We want to close this section with some comments about the localized
anomalies we have just found. The gauge fields $A_{\mu}^{a_f}$, generating
the gauge group $\mathcal{H}_f$, are the only ones that do not vanish
at $y=y_f$.  
We have just seen that localized anomalies, either from
bulk or localized fermions, at the fixed point $y_f$, contain the
factor $a_f(x,y)\,\delta(y-y_f)$ where
\begin{equation}
\label{anom2}
a_f(x,y)=d^{\, a_f b_f c_f}\, \xi^{a_f}\,
F_{\mu\nu}^{\, b_f}\,\tilde{F}^{\, c_f\, \mu\nu}
\end{equation}
where $d^{ABC}={\rm str}\left[T^A T^B T^C\right]$. Using the fact
that $\mathcal{H}_f\supseteq \mathcal{H}$, where $\mathcal{H}$ is the
gauge group of the zero modes $A_\mu^{a}$, we can decompose
\begin{equation}
a_f=a_0+\Delta a_f
\label{descomp}
\end{equation}
where
\begin{equation}
\label{anom3}
a_0(x,y)=d^{\, a b c}\, \xi^{a}\, F_{\mu\nu}^{\, b}\,\tilde{F}^{\, c\,
\mu\nu}
\end{equation}
is the term in the anomaly that has a zero mode, while
\begin{equation}
\label{anom4}
\Delta a_f(x,y)=d^{\, a b \hat c}\, \xi^{a}\,
F_{\mu\nu}^{\, b}\,\tilde{F}^{\, \hat c\, \mu\nu}
+d^{\, a \hat b \hat c}\, \xi^{a}\,
F_{\mu\nu}^{\, \hat b}\,\tilde{F}^{\, \hat c\, \mu\nu}+d^{\,\hat a \hat b
\hat c}\, \xi^{\hat a} \,F_{\mu\nu}^{\,\hat b}\,\tilde{F}^{\,\hat c\,
\mu\nu}
\end{equation}
where the corresponding $T^{\,\hat a}$ generators (elements of
$\mathcal G/\mathcal H$) that appear in (\ref{anom4}) are also in the
coset $\mathcal{H}_f/\mathcal{H}$ and the possible non-vanishing
coefficients $d^{\, a b \hat c}$, $d^{a \hat b \hat c}$ and $d^{\hat a
\hat b \hat c}$ depend on the orbifold compactification and the fixed
point itself. In particular, for fixed points such that
$\mathcal{H}_f=\mathcal{H}$, $\Delta a_f=0$. Notice that the
$a_0$--term in the anomaly has a zero mode that corresponds to
diagrams with three zero mode gauge bosons in $\mathcal H$ as external
legs. Since $\mathcal{H}$ is a common subgroup to all fixed points,
$a_0$ is common to all fixed points. On the other hand $\Delta a_f$
does not have (as a composite operator) any zero mode and it
corresponds diagrammatically to triangular diagrams with less than
three zero mode gauge bosons in $\mathcal H$ as external legs. The
corresponding anomalies would spoil the four-dimensional gauge
invariance by non-renormalizable operators in the effective theory. In
particular the terms $d^{ab\hat c}$ and $d^{a\hat b\hat c}$ would
provide a three-loop contribution to the mass of the zero mass of the
unbroken gauge boson $A_\mu^a$. In the effective four-dimensional
theory this would give rise to a contribution to the gauge boson mass
suppressed by powers of the four-dimensional cutoff $M_c\sim 1/R$,
$M_c^{-2}$ and $M_c^{-4}$ respectively.

We will conclude this section with some comments concerning possible
non-trivial boundary conditions for fermion and gauge boson fields. In
fact we have been implicitly assuming in this section that bulk
fermions (and gauge bosons) in~(\ref{anomaly}) are periodic functions
on the covering space of the compact manifold. However, as we have
seen in section~\ref{wilson} one can introduce in the orbifold
structure some non-trivial boundary conditions, known as
Scherk-Schwarz compactification, possibly breaking the gauge
invariance and thus playing a role in the existence and values of
localized anomalies. If the Scherk-Schwarz boundary conditions satisfy
the constraint (\ref{omegacond}) we have shown that they are
equivalent to a Hosotani breaking where some extra-dimensional
components of gauge fields acquire a VEV and where all fields are
periodic in the covering space of the compact manifold. Furthermore we
have proven that under such conditions the gauge structure at the
different fixed points (${\mathcal H}_f$) is unchanged with respect to
the case where no Scherk-Schwarz breaking is introduced and the
anomaly structure that we have just deduced is equally unchanged.
However we have seen a whole class of Scherk-Schwarz boundary
conditions that can be described by discrete Wilson lines [see
Eq.~(\ref{discreteW})] and that are {\it not} equivalent to a Hosotani
breaking. In this case the unbroken gauge group at the fixed point
$y_f$ is $\mathcal H^W_f\neq\mathcal H_f$ and consequently the
corresponding localized anomaly structure will also change. In fact
the theory should be defined on the covering space and the twists
corresponding to non-trivial boundary conditions should be accounted
in the projector $Q_\psi$ in section~\ref{anomalies}. However as we
have proven, $\mathbb Z_N$ orbifolds on tori in the presence of
Wilson lines are equivalent to orbifolds with larger tori (with radii
$N$ times larger) modded out by a bigger orbifold group and no Wilson
lines~\footnote{Some of these typical examples, in particular the case
of $\mathbb Z_2\otimes\mathbb Z_2^\prime$, have been widely worked out
in the recent
literature~\cite{Scrucca:2001eb,Pilo:2002hu,Barbieri:2002ic}.}. In
that case also the formalism of this section applies to the redefined
orbifold structures.

\section{\sc Green-Schwarz local anomaly cancellation}
\label{GS}

It is well known that bosonic $p$--form fields, typically appearing in
supergravity and string theories, can sometimes cancel the anomalies
produced by fermions~\cite{Green:sg}.  This is the case for the
so-called reducible anomalies, whose corresponding anomaly monomial
can be written as a product of traces~\footnote{We adopt here a
notation in terms of forms where all the products are to be read as
wedge products.}
\be
\text{tr~} F^{p+1}\ \text{tr~} F^{D/2-p}\equiv X_{2p+2}X_{D-2p}
\ee
Since $X_{2p+2}$ is exact one can choose appropriate Chern-Simons
forms $\omega_{2p+1}$ such that \cite{anomalies,alvarez}
\be
X_{2p+2}=d\omega_{2p+1},\quad\delta_\xi \omega_{2p+1}=dX^1_{2p}(\xi).
\label{descent}
\ee
The anomaly is then given by~\footnote{The corresponding term arising
from the descent of  $X_{D-2p}$ can be absorbed in a
counterterm of the form $\omega_{2p+1}\omega_{D-2p-1}$.}
\be
X^1_{2p}(\xi)X_{D-2p}.
\label{reducible}
\ee
To cancel it, one introduces the $2p$-form field $C_{2p}$ which is
coupled to the Chern-Simons forms according to
\be
\frac{1}{2}(dC_{2p}-\omega_{2p+1})*(dC_{2p}-\omega_{2p+1})+C_{2p}X_{D-2p}
\ee
Imposing now the transformation of $C_{2p}$ under the gauge symmetry:
\be
\delta_\xi C_{2p}=X_{2p}^1(\xi)
\ee
the field strength and thus the kinetic term are invariant while the
interaction term transforms as
\be
\delta_\xi (C_{2p}X_{D-2p})=X^1_{2p}(\xi)X_{D-2p}
\ee
canceling the anomalous contribution of Eq.~(\ref{reducible}).  This
mechanism is known as Green-Schwarz (GS) cancellation of reducible
anomalies for gauge theories of dimension $D$.

We now want to make some comments about the form of the anomalies
considered in this section. By construction, Eq.~(\ref{descent}), the
anomalies $X_{2p}^1(\xi)$ satisfy the Wess-Zumino consistency
condition~\cite{WZ}
\be
\delta_{\xi_1} X_{2p}^1(\xi_2)-\delta_{\xi_2}X_{2p}^1 (\xi_1)=
 X_{2p}^1([\xi_1,\xi_2])
\ee
and are then called ``consistent'' anomalies. In 4D, the consistent
anomaly reads $X_4^1(\xi)\propto \rm{str}\,\xi\,
d\left[A\left(dA+\frac{1}{2}A^2 \right)\right]$.  The operation
``str'' means that the trace is symmetrized with respect to all its
factors. On the other hand the covariant path-integral method we have
followed in section~\ref{anomalies}, based on decomposition with
respect to eigenfunctions of the operator $\Dslash$, explicitly
violates Bose symmetry among all the vertices since one of them is
singled out. The anomaly obtained in this way transforms covariantly
under gauge transformations and it is therefore called ``covariant''
anomaly $\propto \rm{tr}\, \xi\, F^2$. Covariant
anomalies do not satisfy the Wess-Zumino consistency conditions. Going
from the covariant to the consistent description of the anomalies
amounts to introducing an extra current $\Delta J^M$ and constitutes a
well defined and standard procedure~\cite{Bardeen:1984pm}. It is well
known \cite{alvarez} that one can arrive directly at the consistent
anomalies via Fujikawa's method by regulating the Jacobian with the
eigenvalues of the operator $\,/\!\!\!\partial+i\,/\!\!\!\!A(1\pm
\Gamma^{D+1})/2$, so one can speculate that the generalization to the
orbifold case amounts to using $\,/\!\!\!\partial+i\,/\!\!\!\!AQ(1\pm
\Gamma^{D+1})/2$.

In gauge theories defined in $D>4$ dimensions compactified on
orbifolds there appear anomalies from bulk propagating fermions
localized at the (four-dimensional) fixed points of the orbifold as we
have seen in section~\ref{anomalies}. Green-Schwarz cancellation of
these localized anomalies can still work in specific cases using
appropriate $p$--forms propagating in the bulk.

\subsection{\sc GS mechanism with bulk four-form}
\label{GSmech1}
If localized anomalies do not cancel globally they are generated by
bulk fermion zero modes and must be canceled by localized fermions
propagating at the fixed points. However, if these anomalies cancel
globally, but not locally, they are generated by fermion non-zero
modes and trigger breakdown of gauge invariance by higher-dimensional
operators suppressed by powers of the cutoff of the four-dimensional
theory.

As pointed out in Ref.~\cite{Scrucca:2002is}, by introducing an
appropriate GS four--form in the bulk, one can cancel globally
vanishing fixed point anomalies of the form
\be
{\cal A}=-\!\int  X_{4}^1(\xi)\,\delta ,
\label{localanom}
\ee
where $X_4^1$ is defined by Eq.~(\ref{descent}) from 
\be
X_6=\text{str}\,(T^A T^B T^C)\,F^A\!\wedge F^B\!\wedge F^C.
\label{6form}
\ee
Notice that although $X_4^1$ in Eq.~(\ref{localanom}) is coupled to
the fixed points, it is defined in the bulk and as such the trace in
Eq.~(\ref{6form}) goes over the group $\mathcal G$.  Therefore this
``anomaly inflow''~\cite{callan} does not in general match with the
form of localized anomalies generated by orbifolds
(\ref{extra})-(\ref{anomlocal}). Moreover it identically vanishes for
groups with only anomaly-free representations in four dimensions, as
e.g.~$G_2$, $SO(10)$ or $E_6$. On the other hand it also vanishes for
mixed $U(1)$ gravitational anomalies if the $U(1)$ originates from a
simple group in the bulk.  In the next subsection we will present a
mechanism which can cancel those mixed anomalies.  The
$\delta$--function in Eq.~(\ref{localanom}) picks out the fixed points
where the anomaly is non-vanishing and has to be considered as a
$(D-4)$--form.  The condition that the anomaly vanishes globally is
simply $\int \delta=0$.  The GS Lagrangian
\be
\frac{1}{2}(dC_{4}-\omega_{5})*(dC_{4}-\omega_{5})+ C_{4}\,\delta
\label{GS1}
\ee
can then cancel the corresponding anomaly. In fact, variation of this
Lagrangian immediately gives the anomaly, which can be seen by using
the transformation $\delta_\xi C_4=X_4^1(\xi)$.

One should be a bit more explicit at this point. The anomaly in
(\ref{localanom}) can be generally written as
$$\sum_f c_f X_4^1(\xi)\delta(y-y_f)$$ where $c_f$ are constants, while the
$\delta$--function in (\ref{GS1}) is given by
$$\delta=\sum_f c_f\delta(y-y_f)$$ and satisfies the condition
$\int\delta=\sum_f c_f=0$. Expanding $X_4^1=X_{4,\,0}^1+\Delta X_4^1$,
where $X_{4,\,0}^1$ descends from $X_{6,0}=\text{str}\,(T^a T^b
T^c)\,F^a\wedge F^b\wedge F^c$ and $\Delta X_4^1\equiv
X_4^1-X_{4,\,0}^1$ we obtain for the anomaly
\be
X_{4,\,0}^1\sum_f c_f\delta(y-y_f)+\sum_f c_f \Delta X_4^1
\delta(y-y_f).
\label{descompo}
\ee
The first term in (\ref{descompo}) has a zero mode that corresponds to
a globally vanishing anomaly since it integrates out to zero.  Notice
that the equation of motion for the zero modes of
$C_{4,\,\mu\nu\rho\sigma}$ implies $\int\delta =0$. Therefore this
mechanism is consistent only for globally vanishing zero-mode
anomalies, corresponding to diagrams with three massless mode gauge
bosons as external legs, as we were assuming~\footnote{This is not
unexpected since otherwise upon dimensional reduction it would
constitute a possibility for cancelling irreducible anomalies by a
Green-Schwarz mechanism in four dimensions.}. However the non-zero
modes in the first term and the second term in (\ref{descompo})
trigger the breakdown of gauge invariance in the effective
four-dimensional theory by higher-dimensional (suppressed) operators
corresponding to some massive mode gauge bosons as external legs in
the triangular diagrams. This anomaly is also cancelled by the
corresponding contributions in the GS four--form $C_4$.

As can easily be verified $\delta$ is closed, $d\delta=0$; the fact
that $\int \delta=0$ guarantees that $\delta$ is exact on the torus,
$\delta=d\eta$ (that this is a sufficient condition can be shown by
explicit construction of $\eta$, it certainly is a necessary one by
Stoke's theorem).  Integrating by parts we find~\footnote{Recall the
following manipulation rules for $p$--forms $\Omega_p$ which we use
here and in the following: $d(\Omega_p\Omega_q)=d\Omega_p
\Omega_q+(-)^{p}\Omega_p d\Omega_q$,
$*\!*\Omega_p=(-)^{p(D-p)+1}\Omega_p$ and
$\Omega_p\Omega_q=(-)^{pq}\Omega_q\Omega_p$.} 
\be
\frac{1}{2}(dC_{4}-\omega_{5})*(dC_{4}-\omega_{5})- dC_{4}\eta
\ee
We can now proceed to integrate out $C_5\equiv dC_4$. To take care of the
constraint $dC_5=0$ we introduce for $D\geq 6$ the $(D-6)$--form
$C_{D-6}$ which plays the role of a Lagrange multiplier
\be
\frac{1}{2}(C_5-\omega_{5})*(C_5-\omega_{5})+C_5(dC_{D-6}-\eta )
\ee
The equation of motion for $C_5$ is now
\be
*(C_5-\omega_{5})+(dC_{D-6}-\eta)=0.
\ee
Substituting back we end up with
\be
\frac{1}{2}*(dC_{D-6}-\eta)(dC_{D-6}-\eta)+\omega_5(dC_{D-6}-\eta)
\label{GS2}
\ee
It is clear that the only gauge--violating piece here is the
counterterm $-\omega_5\,\eta$ whose variation again precisely cancels
the anomaly.  The two descriptions, (\ref{GS1}) and (\ref{GS2}), are
completely equivalent.  One could ask the question whether the
introduction of the counterterm $-\eta\,\omega_5$ would be by itself
sufficient for our purposes. The answer in general is negative since
in $D\geq 6$ this term by itself violates translational invariance in
the bulk and only the described mechanism is allowed by the bulk
symmetries, as it is obvious from Eq.~(\ref{GS1}).

Finally, let us comment on the special cases $D=5,6$. A four--form in
$D=5$ has no physical degrees of freedom due to gauge invariance. In
fact, despite the presence of a kinetic term for $C_4$, there is no
constraint on the field strength which therefore can be integrated out
exactly, leaving over the action
\be
\frac{1}{2}*\!\eta\,\eta-\eta\omega_5
\ee
The zero-form $\eta$ is a simple periodic step-function whose square
term contributes an infinite irrelevant constant to the action. We end
up with the Chern-Simons term
$-\eta\,\omega_5$~\cite{Arkani-Hamed:2001is,Pilo:2002hu,
Barbieri:2002ic,Kim:2002ab,
GrootNibbelink:2002qp,GrootNibbelink:2003gb} whose variation cancels
the anomaly (\ref{localanom}).  On the other hand, in $D=6$, a four--form
has one physical degree of freedom, corresponding to an axion in
Eq.~(\ref{GS2}). One can further clarify the physical picture by
taking the compactification limit where all heavy KK modes
decouple. We expect that the gauge violating counterterms $C_4\delta$
and $\eta\omega_5$ should disappear, since we are considering globally
vanishing anomalies. By dimensionally reducing Eq.~(\ref{GS1}) or
(\ref{GS2}) we find the following low energy actions:
\be \frac{1}{2}(dC-\omega_{3,\,2})*(dC-\omega_{3,\,2})
\label{GS1a}
\ee
\be \frac{1}{2}*\!da\, da+a X_{4,2}
\label{GS2a}
\ee
Here $C_{\mu\nu}$ and $a$ are the zero modes of $(C_4)_{\mu\nu56}$ and
$C_0$ respectively and we have defined the 4D forms
\be
(X_{4,2})_{\mu\nu\rho\sigma}=(X_6)^{(0)}_{\mu\nu\rho\sigma56}=
\text{tr}{F_{[\mu\nu}F_{\rho\sigma}F_{56]}}
\label{someform}
\ee \be
(\omega_{3,2})_{\mu\nu\rho}=(\omega_5)^{(0)}_{\mu\nu\rho56},\quad
d\omega_{3,2}=X_{4,2} 
\label{someocsform}
\ee
where the various $F$-factors are projected over their zero modes.
The actions (\ref{GS1a}) and (\ref{GS2a}) are in fact related by
four-dimensional Poincar\'e duality. The counterterms have indeed
disappeared and the actions are now invariant under four-dimensional
gauge transformations. Nevertheless, in contrast to the $D=5$ case
there is some remnant of the mechanism in the low energy effective
action provided by the non-renormalizable coupling of the axion to
$\text{tr}\{ F^3 \}$.  In fact the existence of zero modes for $C_0$,
$(C_4)_{\mu\nu 56}$ and $\text{tr}\{ F^3 \}$ is a model dependent
question.  For instance in the case of the $\mathbb Z_N$ orbifolds
(see section~\ref{examples6D}) the fact that we deal with orthogonal
transformations
\be
(P_k)_5^5(P_k)_6^6-(P_k)_5^6(P_k)_6^5=1,
\label{det1}
\ee
implies that $(C_4)_{56\mu\nu}$ is left invariant by the orbifold
group and thus has a zero mode. Consequently, its dual $C_0$ has a
zero mode. Eq.~(\ref{det1}) also implies that $\text{tr}\{ F^3 \}$ is
left invariant, but as a composite operator it does not necessarily
have a zero mode.  However, for instance in the case of the $\mathbb
Z_2$ orbifolds (see section~\ref{examples6D} and Ref.~\cite{T2Z2})
with $\mathcal G\to \mathcal H$ orbifold breaking there exists in the
effective theory the remnant non-renormalizable axionic coupling
\be\label{axioncoup}
-\frac{a}{6!\Lambda^3}\text{tr}\,\left\{3\,[A_5,A_6]F_{\mu\nu}\widetilde
F^{\mu\nu}+12\,\partial_\mu A_5 \partial_\nu A_6\widetilde
F^{\mu\nu}\right\}\   
\ee
where we have now rescaled all fields to their canonical dimensions,
$\Lambda$ is the cutoff of the higher-dimensional theory,
$A_i=A_i^{\hat a}T^{\hat a}$ and $F_{\mu\nu}=F^a_{\mu\nu}T^a$.

\subsection{\sc GS mechanism with bulk two--form}
\label{GSmech2}
A mixed $U(1)$ localized anomaly, globally vanishing or not, can be
canceled by a bulk two--form~\cite{GrootNibbelink:2003gb} by a
variation of the mechanism employed in~\cite{Horava:1996ma} to cancel
localized anomalies in 11D M--theory. This mechanism is particularly
interesting for $U(1)$'s that are subgroups of a simple bulk gauge
group $\mathcal G$, for which the previously described mechanism does
not work. Denoting the $U(1)$ gauge bosons at $y_f$ by
$A_\mu^{\alpha_f}$, the mixed anomaly which can be cancelled in this way
is parametrized as
\be
\mathcal A=-\!\int X_4 \sum_f c_f \xi^{\alpha_f}\delta_f  
\label{mixedanomaly}
\ee
where $\delta_f\equiv\delta(y-y_f)$ has again to be considered as a
$(D-4)$--form and $X_4$ is a four-form defined in the bulk by
\be
X_4=\text{str}\,(T^A T^B)\,F^A\!\wedge F^B
\ee
and a similar term involving the curvature two-form $R$.  As in the
case of the mechanism studied in section \ref{GSmech1},
Eq.~(\ref{mixedanomaly}) does not correspond to the general form of
localized $U(1)$ mixed anomalies contributed from bulk and
brane-fermions. When $X_4$ is evaluated on the brane, it splits into a
sum of traces over the subgroups of each simple group within $\mathcal
H_f$, according to the precise branching rule of the fundamental of
$\mathcal G$. To apply the present mechanism, one must ensure that the
anomaly produced by the fermions is of this precise form, which is a
notrivial constraint. Consider now the following GS--Lagrangian:
\be
\frac{1}{2}(dC_2-\omega_3)*(dC_2-\omega_3)-(dC_2-\omega_3)\omega_{D-3}
\label{gs2form}
\ee
where $\omega_{D-3}=\sum_f c_f A^{\alpha_f}\delta_f$. Using that the
field strength $(dC_2-\omega_3)$ is gauge invariant, the gauge
variation of Eq.~(\ref{gs2form}) cancels precisely the anomaly
(\ref{mixedanomaly}) after a partial integration (note that
$d\delta=0$).

As in the previous section, it is illustrative to consider the dual
descrition in terms of a $(D-4)$--form. The corresponding Lagrangian
reads:
\be
\frac{1}{2}*(dC_{D-4}-\omega_{D-3})(dC_{D-4}-\omega_{D-3})+X_4 C_{D-4}.
\label{gsXform}
\ee
The gauge transformation of $C_{D-4}$ is given by
\be
\delta_\xi C_{D-4}=\sum_f c_f\,\xi^{\alpha_f}\,\delta_f,
\ee
so variation of Eq.~(\ref{gsXform}) leads to the same anomaly of
Eq.~(\ref{mixedanomaly}). This mechanism has interesting consequences
if the $U(1)$ gauge boson at the different fixed points has a zero
mode and thus forms a $U(1)$ factor of $\mathcal H$, i.e.~if
$\alpha_f=\alpha\ \forall f$. In this case one has two possibilities
according to whether $c\equiv\sum_f c_f$ vanishes or not: a globally
vanishing and a globally non-vanishing anomaly.  Let us consider the
case of the 5D compactification $S^1/{\mathbb Z}_2$ as an example. We
deal with a two--form or its dual, a one--form. In 5D supergravity
theory the one--form can be identified with the graviphoton.  Coupling
of supergravity to super--YM theory then requires the modification of
the Bianchi-identity by $\delta$ like terms~\cite{Horava:1996ma} which
leads to a field strength as in Eq.~(\ref{gsXform}). The parities for
the $C_1$ and $C_2$ forms are to be inferred from the Lagrangians,
Eq.~(\ref{gs2form}) and (\ref{gsXform}). One finds that the objects
with positive parity are $(C_2)_{\mu\nu}$ and $(C_1)_5$.  Denoting
their corresponding zero modes with $C$ and $a$ respectively, we find
the two actions (dual to each other in 4D)
\be
\frac{1}{2}(dC-\omega_3)*(dC-\omega_3)-c\,(dC-\omega_3)A^\alpha
\label{lowenergy1}
\ee
\be
\frac{1}{2}*(da-c A^\alpha)(da-c A^{\alpha})
+a X_4 
\label{lowenergy2}
\ee
together with the 4D gauge transformations $\delta_\xi
(dC_2-\omega_3)=0$ and $\delta_\xi a=c\,\xi^\alpha$.  Here the forms
$\omega_3$ and $X_4$ carry 4D Lorentz indices only. Now for globally
non-vanishing anomalies ($c\neq0$), this is the usual GS-mechanism in
4D where gauge invariance is restored by the introduction of an axion.
However, this $U(1)$ is spontaneously broken as the axion can be
gauged away and $A_\mu^\alpha$ becomes massive.  On the other hand,
for globally vanishing anomalies, $c=0$, we find a gauge invariant
axion, as expected. Although there is a propagating axion, the gauge
symmetry is not spontaneously broken, contrary to the case when
localized (twisted) axions are used to cancel the globally vanishing
localized anomaly~\cite{Scrucca:2002is}.  Finally let us note that in
case the $U(1)$ gauge boson does not have a zero mode, the anomaly
does not have a zero mode either and the low energy actions are given
by Eq.~(\ref{lowenergy1}) and (\ref{lowenergy2}) with $A_\mu=0$, which
are identical to the $c=0$ case discussed above.

Going to higher dimensions will in general introduce more bosonic
degrees of freedom in the low energy theory, depending on the details
of the compactification.  In 6D the dual of the two--form is again a
two--form $C'_2$. Compactifying on $T^2/{\mathbb Z_N}$ we
obtain zero modes for $(C_2)_{\mu\nu}$ and $(C_2)_{56}$ as well
as $(C'_2)_{\mu\nu}$ and $(C'_2)_{56}$. Denoting the latter by $C'$
and $a'$, we get at low energy from Eq.~(\ref{gsXform}):
\be \frac{1}{2}*(da'-c A^\alpha)(d a'-c
A^{\alpha}) +a' X_4 + \frac{1}{2}*dC'dC'
+C'X_{2,\,2}
\label{lowenergy3}
\ee
Here we defined in analogy to Eq.~(\ref{someform}) the 4D two--form
\be
(X_{2,\, 2})_{\mu\nu}\equiv (X_4)^{(0)}_{\mu\nu 56}
=\text{tr}\,F_{[\mu\nu} F_{56]}.
\ee
The additional terms do not contribute to any possible zero mode
anomaly but leave some non-renormalizable interactions.

\section{\sc $\mathbb{Z}_N$-orbifolds in five and six dimensions}
\label{examples}

In this section we will explicitly consider five and six-dimensional
orbifolds with extra dimensions compactified on the circle $S^1$ and
torus $T^2$, respectively, modded out by the discrete group
$\mathbb{Z}_N$ consistent with the crystallographic action of the
orbifold.

\subsection{\sc 5D orbifolds: $S^1/{\mathbb Z}_2$}
\label{examples5D}
The circle $S^1$ can only be modded out by the discrete group
$\mathbb{Z}_2$.  The $\mathbb{Z}_2$ group consists of the two elements
$\{1,P\}$, $P$ being the reflection $y\rightarrow-y$. The fixed points
of $P$ are $\{0,\pi\}$.  The transformation on the fermions can be
taken as ${\mathcal P}_\frac{1}{2} =\gamma^5$, which implies that
\be
Q_\psi -\Gamma^0Q_\psi\Gamma^0=\lambda\otimes\gamma^5\otimes\hat P
\ee
and therefore
\begin{eqnarray}
{\cal A}&=&-\lim_{M\rightarrow\infty}\text{tr}~ \xi
\left( \lambda\otimes\gamma^5\otimes\hat P\right) 
\exp(-\Dslash^2/M^2)\nonumber\\
&=&-\frac{g^2}{64 \pi^2 }
 \int d^Dx~\xi^AF^B_{\mu\nu}F^C_{\rho\sigma}
\epsilon^{\mu\nu\rho\sigma}\nonumber\\
&& \text{tr}~T^A\lambda\,T^BT^C
\left[ \delta(y)+\delta(y-\pi) \right]
\label{anom5}
\end{eqnarray}

In this simple case the two fixed points $\{0,\pi\}$ are left
invariant by the whole orbifold group $\mathbb{Z}_2$ which means that
at both points the unbroken gauge group coincides with the zero mode
gauge group $\mathcal{H}$ generated by
$\{T^{a}~|~[T^{a},\lambda]=0\}$.  A corresponding irreducible
representation $R_{\mathcal G}$ splits into a direct sum of
irreducible representations $\oplus_i R_{\mathcal{H}}^i$. Since
$\lambda$ commutes with all generators of ${\mathcal H}$, we have that
on each representation space labeled by $i$, $\lambda$ is proportional
to the identity and thus just a number $\lambda_i=\pm$. It can be
taken out of the trace and the contribution of each
$R_{\mathcal{H}}^i$ to the anomaly is easily evaluated as
\begin{equation}
{\cal A}=\mp\frac{g^2}{32 \pi^2 }
 \int d^5 x~\xi^aF^b_{\mu\nu}F^c_{\rho\sigma}
\epsilon^{\mu\nu\rho\sigma}
\text{str}_i~T^a T^b T^c \frac{1}{2}
\left[ \delta(y)+\delta(y-\pi) \right]
\label{anom5bis}
\end{equation}
In this particularly simple case matter in the bulk can only produce
global anomalies. These anomalies can be cancelled either by bulk
fermions [in which case the global coefficient in front of
(\ref{anom5}) would vanish] or by localized fermions at the fixed
points. The latter fermions would generate local anomalies that should
be cancelled by the Green-Schwarz mechanims described in
section~\ref{GS} or equivalently by the introduction of a Chern-Simons
counterterm~\cite{Scrucca:2001eb}.
\subsection{\sc 6D orbifolds: $T^2/{\mathbb Z}_N$} 
\label{examples6D}
The ${\mathbb Z}_N$--group consists of $N$ elements, $\{P^n\}$,
$0\leq n<N$. $P$ is any $N^{th}$ root of unity and the complex
coordinate $z$ of the torus transforms as $z\rightarrow P^n z$.  The
transformation on the fermions can be taken as
\begin{align}
( {\mathcal P}_\frac{1}{2} )^n& =\exp{\left( \frac{i\pi n}{N}
\right)} \exp{\left(\frac{\pi n}{N}\Gamma_5\Gamma_6\right)}\nonumber\\
%& =\exp(i\pi n /N)\left(
%\begin{matrix}\exp(-i\pi n /N)&0\\0&\exp(i\pi n /N)\end{matrix} \right)\\
&=\cos\left( \frac{\pi
n}{N} \right)\exp\left( \frac{i\pi n}{N} \right) +\sin\left( \frac{\pi
n}{N} \right)\exp\left(\frac{i\pi n}{N} \right)\Gamma_5\Gamma_6
\label{Pfermion}
\end{align}
Here, in addition to a pure Lorentz rotation we have allowed for an
additional phase in order to have ${\mathcal
P}_{\frac{1}{2}}^N=+1$. This can be interpreted as a global phase
rotation, which is always allowed since in 6D the fermions are complex
and thus there is a global $U(1)$ fermion-number symmetry. If we also
have a chiral $U(1)$ we can instead allow for a similar factor
\be
\exp \left( \frac{i\pi n(\alpha+\beta\Gamma^7)}{N}\right)
\label{alphabeta}
\ee
where $\alpha$ and $\beta$ are
integers with $\alpha+\beta=$~odd.

The crystallographic principle only allows for the four values
$N=2,3,4,6$~\cite{book}.  The corresponding geometries are displayed
in Figs.~\ref{z2z4fig} and~\ref{z3z6fig} in appendix~B.  To denote the
fixed points, we define the lattice by the vectors ${1,\theta}$, where
$\theta=i$ in the case of ${\mathbb Z_2}$ and ${\mathbb Z_4}$, and
$\theta=\exp(2\pi i/3)$ in the cases of ${\mathbb Z_3}$ and ${\mathbb
Z_6}$.  We then define
\begin{equation}
z_{ab}=\left\{\begin{array}{ll}
0&\text{for }(a,b)=(0,0)\\
(a+b\,\theta)/2&\text{for }(a,b)=(1,0),(0,1),(1,1)\\
(a+b\,\theta)/3&\text{for }(a,b)=(2,1),(1,2)
\end{array}
\right.
\label{fp}
\end{equation}
\begin{itemize}
\item $N=2$.  The four fixed points of $P$ are
$\{z_{00},z_{10},z_{01},z_{11}\}$: $\mathbb{G}_f=\mathbb{Z}_2$,
$\forall f$. This is an example where $\mathcal{H}_f=\mathcal{H}$,
$\forall f$.
\item $N=3$.  The three common fixed points of $P$ and $P^2$ are
$\{z_{00},z_{21},z_{12}\}$: $\mathbb{G}_f=\mathbb{Z}_3$, $\forall
f$. This is another example where $\mathcal{H}_f=\mathcal{H}$,
$\forall f$.
\item $N=4$.  The two common fixed points of $P$ and $P^3$ are
$\{z_{00},z_{11}\}$,
whereas $P^2$ has the four fixed points,
$\{z_{00},z_{10},z_{01},z_{11}\}$:
$\mathbb{G}_{00}=\mathbb{G}_{11}=\mathbb{Z}_4$,
$\mathbb{G}_{10}=\mathbb{G}_{01}=\mathbb{Z}_2$. Here
$\mathcal{H}_{00}=\mathcal{H}_{11}=\mathcal{H}$ while
$\mathcal{H}_{10}=\mathcal{H}_{01}$ can be a larger subgroup.

\item $N=6$.  The three common fixed points of $P^2$ and $P^4$ are
$\{z_{00},z_{21},z_{12}\}$.  The elements $P$ and $P^5$ leave only
$z_{00}$ invariant. Finally $P^3$ leaves four points invariant:
$\{z_{00},z_{10},z_{01},z_{11}\}$.  $\mathbb{G}_{00}=\mathbb{Z}_6$,
$\mathbb{G}_{10}=\mathbb{G}_{01}=\mathbb{G}_{11}=\mathbb{Z}_2$,
$\mathbb{G}_{12}=\mathbb{G}_{21}=\mathbb{Z}_3$. Here the only fixed
point with gauge group $\mathcal{H}$ is the origin while all the
others can have larger gauge groups realized by gauge bosons that do
not have zero modes and
$\mathcal{H}_{10}=\mathcal{H}_{01}=\mathcal{H}_{11}$,
$\mathcal{H}_{12}=\mathcal{H}_{21}$.
\end{itemize}
Note that Lefschetz' formula gives the number of fixed points
correctly in each case, i.e.~$\nu(n,N)=4\sin^2(\pi n/N)$.

The {$T^2/{\mathbb Z}_N$} orbifold thus provide the anomalies:
\begin{eqnarray}\label{anomzetaN}
{\cal A}&=&\mp\lim_{M\rightarrow\infty}\text{tr}~ \xi\, Q_\psi
\Gamma^{7}\exp(-\Dslash^2/M^2)\nonumber\\
&=&\mp\frac{g^3}{384 \pi^3 N}\epsilon^{MNRSTU}
\int d^6x~\xi^A F^B_{MN}F^C_{RS}F^D_{TU}~
\text{tr}~T^AT^BT^CT^D\nonumber\\
&&\pm\frac{ig^2}{16 \pi^2 N}\epsilon^{\mu\nu\rho\sigma}
 \int d^6x~\xi^AF^B_{\mu\nu}F^C_{\rho\sigma}
 \sum_{n=1}^{N-1} \text{tr}~T^A\lambda^nT^BT^C
\frac{\exp\left(\frac{i\pi n}{N} \right)}{4\sin\left( \frac{\pi
n}{N} \right)}\ \delta_n(z)
\end{eqnarray}
Notice that the first term in (\ref{anomzetaN}) is the usual
six-dimensional bulk anomaly~\cite{6Danomalies} that needs to be
canceled and constrains the bulk matter content of the theory. The
second term gives rise to localized anomalies at the orbifold fixed
points.  Here we have called $\delta_n(z)$ the sum of delta functions
picking out the fixed points of the corresponding element $P^n$ as
listed above.

The gauge group ${\mathcal G}$ generated by $\{T^A\}$ breaks down at
the fixed points left invariant by the whole orbifold group
$\mathbb{Z}_N$ (including the origin) to a subgroup ${\mathcal H}$
generated by $\{T^{a}~|~[T^{a},\lambda]=0\}$. In particular ${\mathcal
H}$ is the group of zero modes. A corresponding irreducible
representation $R_{\mathcal G}$ splits into a direct sum of
irreducible representations $\oplus_i R_{\mathcal{H}}^i$. Since
$\lambda$ commutes with all generators of ${\mathcal H}$, we have that
on each irreducible block labeled by $i$, the matrix $\lambda$ is
proportional to the identity
\begin{equation}
\lambda=\left(\begin{array}{ccc}
\lambda_1\Id_{d_1}   &&\\
& \lambda_2\Id_{d_2} &\\
&& \ddots\\ 
\end{array}
\right).
\end{equation}
There the $\lambda_i$ factors are just numbers and can be expressed as
$$\lambda_i=\exp(2\pi i r_i/N),\quad r_i\neq r_j\ (\text{mod}\ N)$$
where the vector $\vec{r}=(r_i)$ defines the symmetry breaking pattern.

At an arbitrary fixed point $z_f$ left invariant by a subgroup
$\mathbb{Z}_{N_f}\subseteq\mathbb{Z}_{N}$, $N_f\leq N$, with generator
$P^{n_f}$, $n_f=N/N_f$ the gauge group $\mathcal{G}$ breaks down to
the subgroup $\mathcal{H}_f$ generated by
$\{T^{a_f}~|~[T^{a_f},\lambda^{n_f}]=0\}$ and the irreducible
representation $R_{\mathcal G}$ splits into a direct sum of
irreducible representations $\oplus_i R_{\mathcal{H}_f}^i$. Since
$\lambda^{n_f}$ commutes with all generators of ${\mathcal H}_f$, we
have that on each representation space labeled by $i$,
\begin{equation}\label{condicion}
\lambda^{n_f}=\left(\begin{array}{ccc}
\exp\left(2\pi i\frac{r_1}{N_f}\right)\,\Id_{d_1}   &&\\
& \exp\left(2\pi i\frac{r_2}{N_f}\right)\,\Id_{d_2} &\\
&& \ddots \\ 
\end{array}
\right),\quad r_i\neq r_j\ (\text{mod}\ N_f)\ .
\end{equation}
Again the vector $\vec{r}$, subject to the condition
(\ref{condicion}), defines the symmetry breaking pattern
$\mathcal{G}\to\mathcal{H}_f$. In this way $\lambda^{n_f}$ can be
taken out of the trace and the contribution of each
$R_{\mathcal{H}_f}^i$ to the localized anomaly is easily
evaluated. The result is
\be \mp\frac{g^2}{32
\pi^2}\epsilon^{\mu\nu\rho\sigma}\sum_f\sum_{r_i}\int d^6x~\xi^{\,a_f}~
F^{\,b_f}_{\mu\nu}~F^{\,c_f}_{\rho\sigma}~\text{str}_i~T^{\,a_f}T^{\,b_f}T^{\,c_f}
\sigma_f(r_i)\ \delta(z-z_f) ,
\label{braneanomaly}
\ee
where the sum is extended to the different values of $r_i$
$(\text{mod}\ N_f)$, the trace is to be taken in the representation
$R_{\mathcal{ H}_f}^i$ and $\sigma_f$ depends only on the value of
$r_i$ defining the symmetry breaking pattern at the corresponding
fixed point.  Note that we have not demanded $\mathcal H_f$ to be
simple, so Eq.~(\ref{braneanomaly}) contains all possible pure
gauge-anomalies, i.e.~non-abelian, abelian and mixed ones.  For
reference we list $\sigma_f(r)$ in table~\ref{sigmavalues}.

\begin{table}[htb]
\begin{center}
\begin{tabular}{||cc|cccccc|c||}\hline
$N$&$r$&$z_{10}$&$z_{01}$&$z_{11}$&$z_{00}$&$z_{21}$&$z_{12}$&$\Sigma$\\
\hline
2&0&$+\myfrac{1}{4}$&$+\myfrac{1}{4}$&$+\myfrac{1}{4}$&$+\myfrac{1}{4}$
&-&-&$+1$\\
&1&$-\myfrac{1}{4}$&$-\myfrac{1}{4}$&$-\myfrac{1}{4}$&$-\myfrac{1}{4}$
&-&-&$-1$\\
\hline
3&0&-&-&-&$+\myfrac{1}{3}$&$+\myfrac{1}{3}$&$+\myfrac{1}{3}$&$+1$\\
&1&-&-&-&$\phantom{-}0$&$\phantom{-}0$&$\phantom{-}0$&$\phantom{-}0$\\
&2&-&-&-&$-\myfrac{1}{3}$&$-\myfrac{1}{3}$&$-\myfrac{1}{3}$&$-1$\\
\hline
4&0&$+\myfrac{1}{8}$&$+\myfrac{1}{8}$&$+\myfrac{3}{8}$&$+\myfrac{3}{8}$
&-&-&$+1$\\
&1&$-\myfrac{1}{8}$&$-\myfrac{1}{8}$&$+\myfrac{1}{8}$&$+\myfrac{1}{8}$
&-&-&$\phantom{-}0$\\
&2&$+\myfrac{1}{8}$&$+\myfrac{1}{8}$&$-\myfrac{1}{8}$&$-\myfrac{1}{8}$
&-&-&$\phantom{-}0$\\
&3&$-\myfrac{1}{8}$&$-\myfrac{1}{8}$&$-\myfrac{3}{8}$&$-\myfrac{3}{8}$
&-&-&$-1$\\
\hline
6&0&$+\myfrac{1}{12}$&$+\myfrac{1}{12}$&$+\myfrac{1}{12}$&$+\myfrac{5}{12}$
&$+\myfrac{1}{6}$&$+\myfrac{1}{6}$&$+1$\\
&1&$-\myfrac{1}{12}$&$-\myfrac{1}{12}$&$-\myfrac{1}{12}$&$+\myfrac{1}{4}$
&$\phantom{-}0$&$\phantom{-}0$&$\phantom{-}0$\\
&2&$+\myfrac{1}{12}$&$+\myfrac{1}{12}$&$+\myfrac{1}{12}$&$+\myfrac{1}{12}$
&$-\myfrac{1}{6}$&$-\myfrac{1}{6}$&$\phantom{-}0$\\
&3&$-\myfrac{1}{12}$&$-\myfrac{1}{12}$&$-\myfrac{1}{12}$&$-\myfrac{1}{12}$
&$+\myfrac{1}{6}$&$+\myfrac{1}{6}$&$\phantom{-}0$\\
&4&$+\myfrac{1}{12}$&$+\myfrac{1}{12}$&$+\myfrac{1}{12}$&$-\myfrac{1}{4}$
&$\phantom{-}0$&$\phantom{-}0$&$\phantom{-}0$\\
&5&$-\myfrac{1}{12}$&$-\myfrac{1}{12}$&$-\myfrac{1}{12}$&$-\myfrac{5}{12}$
&$-\myfrac{1}{6}$&$-\myfrac{1}{6}$&$-1$\\
\hline
\end{tabular}
\label{sigmavalues}
\caption{\it The values of $\sigma_f$ appearing in
Eq.~(\ref{braneanomaly}) for the different ${\mathbb Z}_N$ orbifolds.
In the last column we give the sum of all the contributions at the
different fixed points.  }
\end{center}
\end{table}

Notice that, as we have stressed, different fixed points $z_f$ have
different unbroken groups $\mathcal{H}_f\supseteq\mathcal{H}$ all of
them having the common subgroup of zero modes $\mathcal{H}$. This
means that the anomaly with coefficients given in
table~\ref{sigmavalues} does in general not vanish after integration
of the extra dimensions. However by decomposing the anomaly at the
fixed point $z_f$ (\ref{braneanomaly}) with respect to the generators
of $\mathcal{H}\oplus \mathcal{H}_f/\mathcal{H}$ as in (\ref{descomp})
we obtain
\begin{equation}
\label{descomp2}
\mathcal{A}_f=\mathcal{A}_0 +\Delta\mathcal{A}_f
\end{equation}
where the term $\mathcal{A}_0$ includes the generators of
$\mathcal{H}$ for all fixed points. The corresponding anomaly contains
a zero mode that vanishes globally for the cases that sum up to zero
in the last column of table~\ref{sigmavalues}. 
We see that only in the cases $r=0$ or $r=N-1$ there are globally
non-vanishing anomalies.  It is a simple matter to verify that
$\lambda_i{\cal P}_\frac{1}{2}$ has only eigenvalues $+1$ for
$r_i=0$\ ($r_i=N-1$), in which case there is a single zero mode
left-handed (right handed) 4D Weyl fermion in the represetation
$\mathcal R_{\mathcal H}^i$.
On the other hand the anomaly corresponding to
$\sum_f\Delta\mathcal{A}_f$ includes the generators of
$\mathcal{H}_f/\mathcal{H}$ and contains no zero mode.

Let us compare this form of the anomaly with the ones from brane
fermions and bulk GS-forms. The contribution from a localized fermion
is obtained by setting $\sigma_f=1$ in Eq.~(\ref{braneanomaly}).  The
sum is of course over all representations appearing at $y_f$.  The
contribution from the GS four-form (after conversion from cosistent to
covariant anomaly) is obtained by setting $\sigma_f(r_i)=c_f$, where
$c_f$ are arbitrary coefficients independent of $i$ summing up to
zero. The sum over representations is fixed by the branching rule of
the fundamental of $\mathcal G$, i.e.~$\mathcal G=\oplus_i \mathcal
R_{\mathcal H_f}^i$. 
Finally, the contribution from the GS two-form is
given by
\be \frac{g^2}{32
\pi^2}\epsilon^{\mu\nu\rho\sigma}\sum_f\sum_{i}\int d^6x~\xi^{\,\alpha_f}\,
F^{\,b_f}_{\mu\nu}~F^{\,c_f}_{\rho\sigma}~\text{str}_i~T^{\,b_f}T^{\,c_f}
c_f\ \delta(z-z_f),
\ee
where $\alpha_f$ labels the $U(1)$ subgroup of $\mathcal G$ and the
sum over representatinos is again given by the branching rule. Notice
that as described in section \ref{GS} the condition for $U(1)$ not to
be spontaneously broken is $\sum c_f=0$. It should be clear that there
is no general recipe to achieve anomaly cancellation for a general
breaking on a 6D-orbifold. Instead we will give a simple example in
the next subsection.

\subsection{\sc Examples}

This subsection is devoted to illustrate the previous methods with a
simple but instructive example based on $T^2/{\mathbb Z}_4$.  Let us
therefore consider $\mathcal{G}= SU(3)$ in the bulk and apply the
inner automorphism $\lambda=\text{diag}(-1,-1,1)$ to break it down to
$\mathcal H=SU(2)\otimes U(1)$. This specific choice gives the
following gauge groups at the branes: $\mathcal H_{00}=\mathcal
H_{11}=SU(2)\otimes U(1)$ and $\mathcal H_{01}=\mathcal H_{10}=SU(3)$.
Let us first ensure 6D anomaly--freedom by choosing the vector like
bulk fermion content $\mathbf 3_{L_6}\oplus\mathbf 3_{R_6}$. Baring
Dirac mass--terms for those fermions, we are free to choose a relative
$\mathbb Z_4$--phase between the two chiral fermions\footnote{In fact
6D chiral symmetry allows for a non-zero $\beta$ in
Eq.~(\ref{alphabeta}), so equivalently one can work with a 6D Dirac
fermion triplet with parity assignment $\lambda\otimes\mathcal
P_{\frac{1}{2}}'$ where $P_{\frac{1}{2}}'$ is defined by $\alpha=0,\
\beta=1$.}, so let us take $\lambda$ as the parity for $\mathbf
3_{L_6}$ and $-i\lambda={\rm diag}(i,i,-i)$ as the parity for $\mathbf
3_{R_6}$.  We have the usual branching ratio $\mathbf 3\rightarrow
\mathbf 2_1\oplus\mathbf 1_{-2}$.  The necessary ingredients for
Eq.~(\ref{braneanomaly}) are given in table \ref{su3example}. The
anomaly is then given by
\begin{table}[htb]
\begin{center}
\begin{tabular}{||l|l|c|c|c||}
\hline
$\mathcal R_{\mathcal G}$&fixed points&$\mathcal R_{\mathcal H_f}$
   &$r$&$\sigma$\\         
\hline
 $\begin{array}{l}\\ \mathbf 3_{L_6}\end{array}$ 
   &$z_{00}$ and $z_{11}$
   &$\begin{array}{c}\mathbf 2_1\\ \mathbf 1_{-2}\end{array}$   
   &$\begin{array}{l}2\\0\end{array}$
   &$\begin{array}{l} \myfrac{-1}{8}  \\ \myfrac{+3}{8}  \end{array}$\\
\cline{2-5}
                  &$z_{10}$ and $z_{01}$&$\mathbf 3$     &2&$\myfrac{+1}{8}$\\
\hline
\hline
 $\begin{array}{l}\\ \mathbf 3_{R_6}\end{array}$ 
   &$z_{00}$ and $z_{11}$
   &$\begin{array}{c}\mathbf 2_1\\ \mathbf 1_{-2}\end{array}$   
   &$\begin{array}{l}1\\3\end{array}$
   &$\begin{array}{l} \myfrac{+1}{8}  \\ \myfrac{-3}{8}  \end{array}$\\
\cline{2-5}
                  &$z_{10}$ and $z_{01}$&$\mathbf 3$     &1&$\myfrac{-1}{8}$\\
\hline
\end{tabular}
\end{center}
\caption{An $SU(3)$ example.}
\label{su3example}
\end{table}
\begin{align}
-\int d^6x
&\left([\delta(z-z_{00})+\delta(z-z_{11})]
\left\{-\frac{1}{4}X_4^{1,\,\text{cov}}(\mathbf 2_1)+
\frac{3}{4}X_4^{1,\,\text{cov}}(\mathbf 1_{-2})\right\}\right.\\
&\left.+[\delta(z-z_{10})+
\delta(z-z_{01})]~\frac{1}{4}X_4^{1,\,\text{cov}}(\mathbf 3)\right) 
\label{su3anomaly}
\end{align}
where we have defined $$X_4^{1,\,\text{cov}}(\mathcal R_{\mathcal
H_f})=\frac{g^2}{32 \pi^2}\text{tr}\,\xi F\widetilde F$$ and the trace
is to be computed in the specified representation. Note that in
Eq.~(\ref{su3anomaly}) the 6D right handed fermion contributes with an
additional minus sign. The 4D zero mode anomaly is immediately read off
\be
-\int d^6x
~2 X_4^{1,\,\text{cov}}(\mathbf 1_{-2}).
\ee
To cure this, we have to add two 4D left handed singlets with charge
$+2$ on the branes. In order to be able to cancel the localized
anomaly, the only solution is to distribute them symmetrically on the
fixed points $z_{00}$ and $z_{11}$ such that the localized anomaly becomes
\begin{align}
-\int d^6x
&\left([\delta(z-z_{00})+\delta(z-z_{11})]
\left\{-\frac{1}{4}X_4^{1,\,\text{cov}}(\mathbf 2_1)
-\frac{1}{4}X_4^{1,\,\text{cov}}(\mathbf 1_{-2})\right\}\right.\\
&\left.+[\delta(z-z_{10})+
\delta(z-z_{01})]~\frac{1}{4}X_4^{1,\,\text{cov}}(\mathbf 3)\right) 
\label{su3anomaly1}
\end{align}
which has the form of the anomaly in Eq.~(\ref{localanom}).  
Indeed, we can now add a Green--Schwarz four--form coupled to the brane in
the following way:
\be
-\int d^6x
\left(\frac{1}{4}[\delta(z-z_{00})+\delta(z-z_{11})]
-\frac{1}{4}[\delta(z-z_{10})+
\delta(z-z_{01})]\right )(C_4)_{\mu\nu\rho\sigma}\epsilon^{\mu\nu\rho\sigma} 
\label{su3anomaly2}
\ee
Using that 
\be
\delta_\xi C_4=X_4^1(\mathbf 3)=\left\{
\begin{array}{ll}
X_4^{1}(\mathbf 2_1)+X_4^{1}(\mathbf 1_{-2})&\text{on } z_{00},\ z_{11}\\
X_4^1(\mathbf 3)&\text{on } z_{10},\ z_{10}
\end{array}\right.
\ee
we see that the variation of Eq.~(\ref{su3anomaly2}) will cancel with
the contribution of Eq.~(\ref{su3anomaly}) upon a straightforward
conversion from covariant to consistent anomalies,
$X_4^{1,\,\text{cov}}\rightarrow X_4^1$.

\section{\sc Conclusion}\label{conclusion}

In this paper we have analyzed the appearance of anomalies in
arbitrary orbifold gauge theories using the path-integral method. We
generically consider orbifolds with periodic fields in the covering
space of the compact manifold and show that the existence of anomalies
localized at the orbifold fixed points is a general feature of
orbifold constructions.  While cancellation of those localized
anomalies imply at the same time that the 4D effective theory is
anomaly free, the converse is not true. In general, localized fermions
have to be introduced to cancel those anomalies.  A further source of
localized anomalies are bosonic $p$-form fields in the bulk, typically
appearing in supergravity and string theories.  We can identify two
class of mechanisms which can be employed to cancel localized
anomalies of a specific form.  A two-form (or its dual a $D-4$ form)
produces mixed $U(1)$ (including $U(1)^3$) anomalies, which can be
globally vanishing or not.  A four-form (or its dual $D-6$ form)
produces all kind of pure gauge anomalies. Those anomalies are always
globally vanishing and the mechanism reduces to a simple Chern-Simons
counterterm in 5D where the four-form is not propagating and can be
integrated out.  If a localized $U(1)$ anomaly is canceled by a GS
two-form in the bulk, the corresponding $U(1)$ in 4D is spontaneously
broken if and only if it is globally non-vanishing.  The form of the
bosonic contributions is quite special from the four-dimensional point
of view and does in general not match the contribution from the
fermions. In particular, the anomaly produced by the four-form is
descended from the anomaly polynomial
\be
\sum_f c_f\, \text{str}\,(T^A T^B T^C)\,F^A\!\wedge 
F^B\!\wedge F^C\wedge\delta_f,
\label{poly}
\ee
with the integrability condition $\int_C \delta=\sum_f c_f=0$, which
is not the general form of localized anomalies generated by fermions
(\ref{extra},\,\ref{anomlocal}). The form of the polynomial Eq.~(\ref{poly}) is
determined by the precise branching rule of the fundamental of
$\mathcal G$ under the breaking $\mathcal G\rightarrow \mathcal
H_f$. In particular, it vanishes for groups $\mathcal G$ with only
anomaly-free representations in four dimensions, as e.g.~$G_2$,
$SO(10)$ or $E_6$. Similarly, the mixed anomaly created by the
two-form descends from the anomaly polynomial
\be
\sum_f c_f \,\text{str}\,(T^A T^B)\,F^A\!\wedge F^B\!\wedge 
F^{\alpha_f}\wedge\delta_f,
\ee
and a similar part proportional to $\text{tr} R^2$. Again, the form of
the created ``anomaly inflow'' is not the most general 4D one produced
by a gauge group $\mathcal H_f$.  To conclude, if the anomalies created
by bulk and brane fermions are not vanishing locally, they must be of
the described form to be canceled by the Green-Schwarz mechanism.
Furthermore, zero mode anomalies of bulk fermions have in most cases
to be canceled by localized fermions, with the possible exception of
$U(1)$ mixed anomalies which imply the spontaneous breakdown of the
$U(1)$~\footnote{Also note the possibility of canceling local mixed
anomalies by localized axions~\cite{Scrucca:2002is}. In this case the $U(1)$ is
spontaneously broken even if there would be no anomaly in the four
dimensional effective theory.}. All this puts very strong constraints on
model building. 

The bosonic states can leave non-renormalizable axionic interactions
in the low energy theory, which in 5D and 6D are given by
Eqs.~(\ref{GS2a}), (\ref{lowenergy2}), and (\ref{lowenergy3}). They
typically involve couplings of zero-forms $a(x)$ and two-forms $C(x)$
like
\be
a\,\text{tr}\,F_{[\mu\nu}F_{\rho\sigma]}\epsilon^{\mu\nu\rho\sigma},\quad
C_{\mu\nu}\text{tr}\,F_{[\rho\sigma}F_{56]}\epsilon^{\mu\nu\rho\sigma},\quad
a\,\text{tr}\,F_{[\mu\nu}F_{\rho\sigma}F_{56]}\epsilon^{\mu\nu\rho\sigma}
\ee
where the zero modes of $F_{\mu\nu}$, $F_{\mu5}$ and $F_{56}$ are
understood.

We also have analyzed orbifolds in the presence of non-trivial
Scherk-Schwarz phases (Wilson lines) on the covering space further
breaking the gauge symmetry by the boundary conditions. We have
established the consistency condition on Wilson lines and their
relation with the Hosotani breaking where an extra-dimensional
component of a gauge field acquires a VEV. We have shown that in five
dimensions the consistency condition of Wilson lines is trivially
fulfilled and that a Hosotani breaking can always be interpreted as a
Scherk-Schwarz breaking. However in $D\geq 6$ the consistency
condition of Wilson lines and the equivalence between the Hosotani and
Scherk-Schwarz breakings are not automatically satisfied and impose
non-trivial constrains. We have proved that if the conditions for the
equivalence between the Scherk-Schwarz and Hosotani breakings are
fulfilled the gauge groups localized at the orbifold fixed points do
not change with respect to the case where fields satisfy trivial
(periodic) boundary conditions. Obviously under those circumstances
the localized anomalies do not change either. There is however a
special case where the Scherk-Schwarz breaking is not equivalent to a
Hosotani breaking: it is the case of discrete Wilson lines defined
along the Cartan subalgebra of the gauge group. In that case the
localized gauge group structure, and consequently the localized
anomalies, get modified with respect to the case with no Wilson lines
and new projections should be introduced in our analysis of localized
anomalies in section~\ref{anomalies}. However as we have observed we
can describe the corresponding theory as one without Wilson lines
defined in a larger compact space modded out by a larger orbifold
group. Then the general analysis done in section~\ref{anomalies}
applies to the new structure.

Finally and to illustrate the previous general ideas we have
explicitly constructed in section~\ref{examples} a class of orbifolds
in five and six dimensions based on the $\mathbb Z_N$ discrete
groups. We have assumed for the different geometries an arbitrary
gauge group invariance and general orbifold automorphisms breaking it
to different subgroups at different fixed points. Using the results in
section~\ref{examples} it would be straightforward to construct
particular field-theoretical models with different gauge structure as
particular applications. We have not tried to present any of those
applications here since they are outside the scope of the present
paper.

\section*{\sc Acknowledgments}
The work of GG was supported by the DAAD.
\vspace{1cm}
\begin{center}
{\sc\Large Appendix}
\end{center}

\appendix

\section{\sc Conventions}

In $D=4$ we use the following representation of $\gamma$ matrices
\be
\label{gamma4}
\gamma^\mu=\left(
\begin{array}{cc}
0&\sigma^\mu\\
\bar\sigma^\mu &0
\end{array}\right)
\ee
where $\sigma^\mu=(1,\vec\sigma)$, $\bar\sigma^\mu=(1,-\vec\sigma)$
and $\vec{\sigma}$ are Pauli matrices. We define
$\gamma^5=diag(1,-1)$.

In $D=6$ we use
\be
\Gamma^\mu=\gamma^\mu\otimes\sigma^3,
\quad\Gamma^5=1\otimes i\sigma^1,
\quad\Gamma^6=1\otimes i\sigma^2,
\quad\Gamma^7=\gamma^5\otimes\sigma^3
\ee
For any even $D$ we apply this formula recursively.  For the trace
over the $\Gamma$ matrices we use the following formula:
\be 
\text{tr}~\Gamma^{D+1}\Gamma^{M_1}\dots\Gamma^{M_r}=
-i(-2i)^r\epsilon^{M_1\dots M_r} 
\ee
where $r=D/2$.

\section{\sc Six-dimensional orbifolds}

\begin{figure}[hbt]
\centering
\epsfig{file=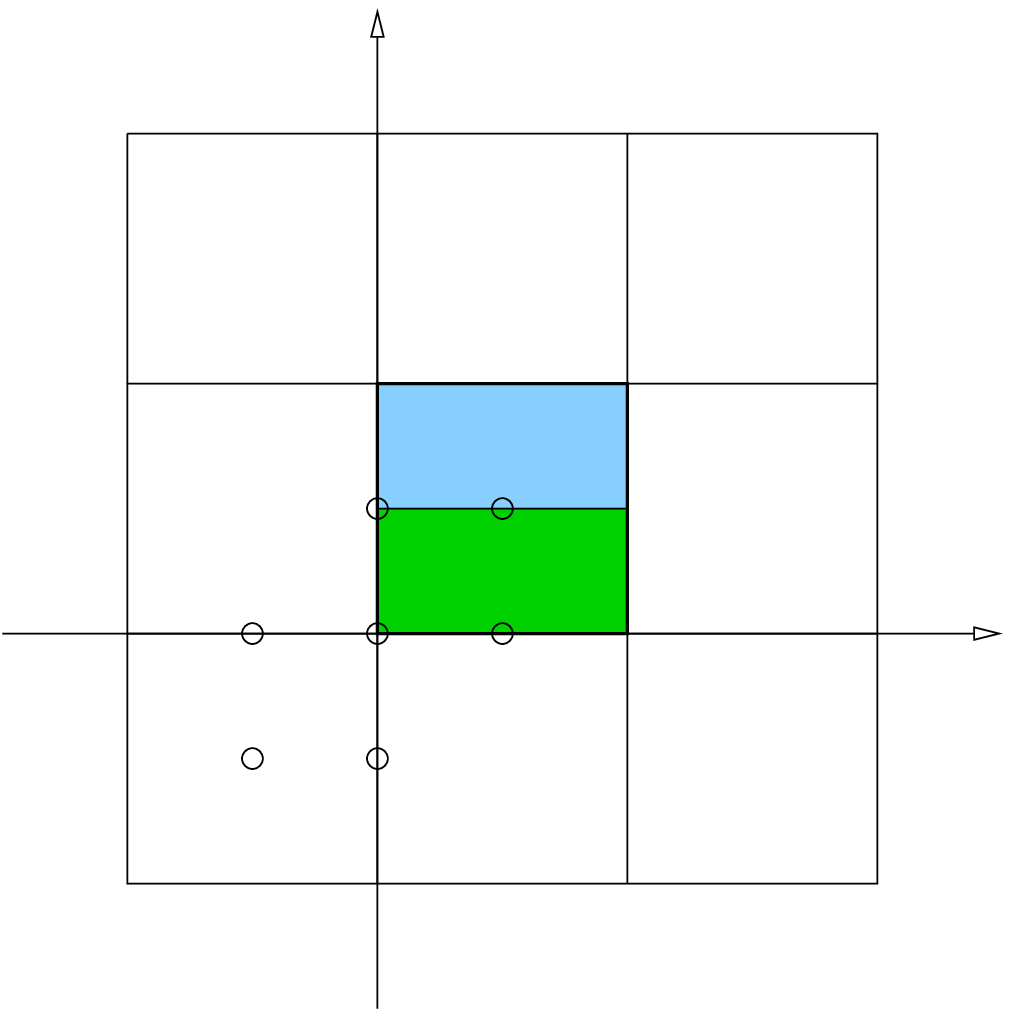,width=.45\linewidth}
\epsfig{file=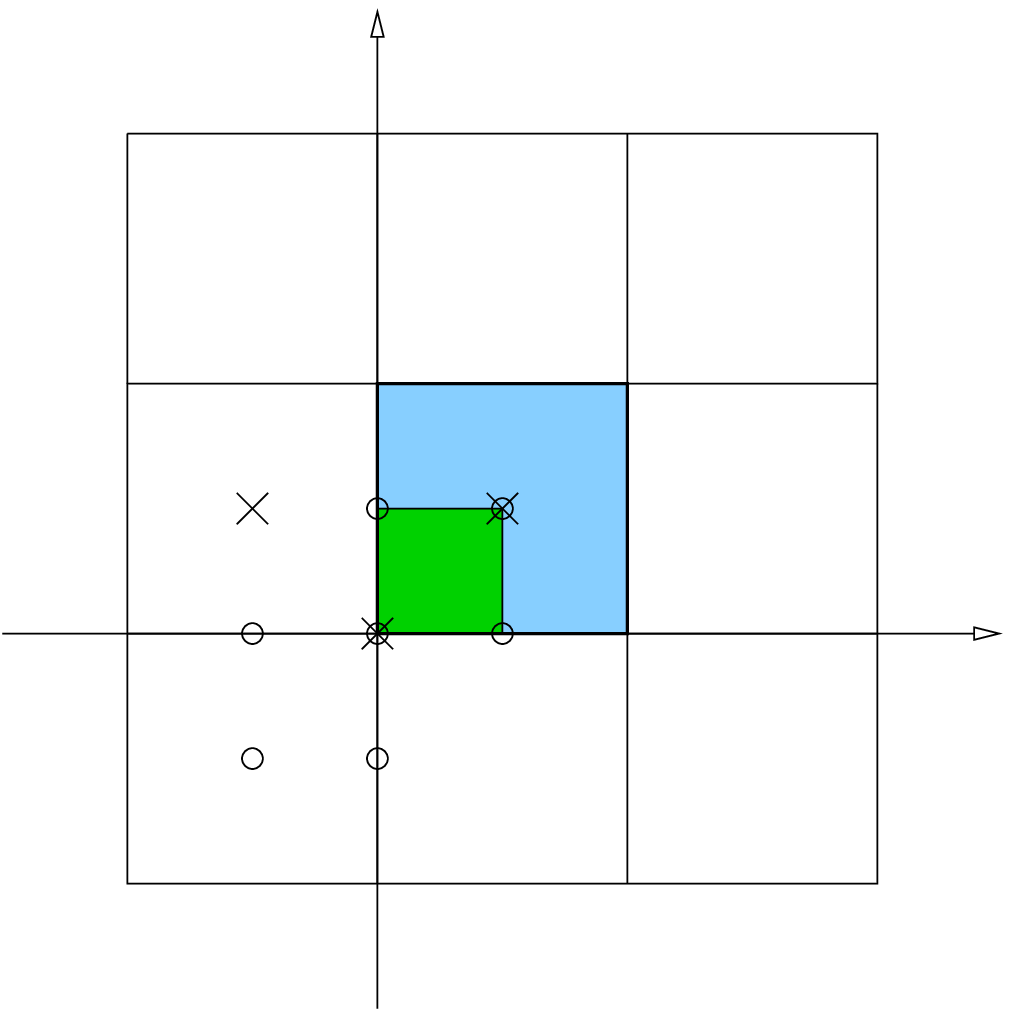,width=.45\linewidth}
\caption{\it $T^2/{\mathbb Z}_2$ (left panel) and $T^2/{\mathbb Z}_4$
(right panel) orbifolds.}
\label{z2z4fig}
\end{figure}

$T^2/{\mathbb Z}_2$ and $T^2/{\mathbb Z}_4$ orbifolds are shown in
Fig.~\ref{z2z4fig}. The fundamental domain of the torus are all shaded
regions while the fundamental domain of the orbifold is only the
darkly shaded (green) one. Open circles correspond to fixed points of
rotations of $\pi$, crosses of rotations of $\pi/2$. The images of the
fixed points are displayed as well and are easily seen to be
equivalent to its sources by a torus shift.

\begin{figure}[htb]
\centering
\epsfig{file=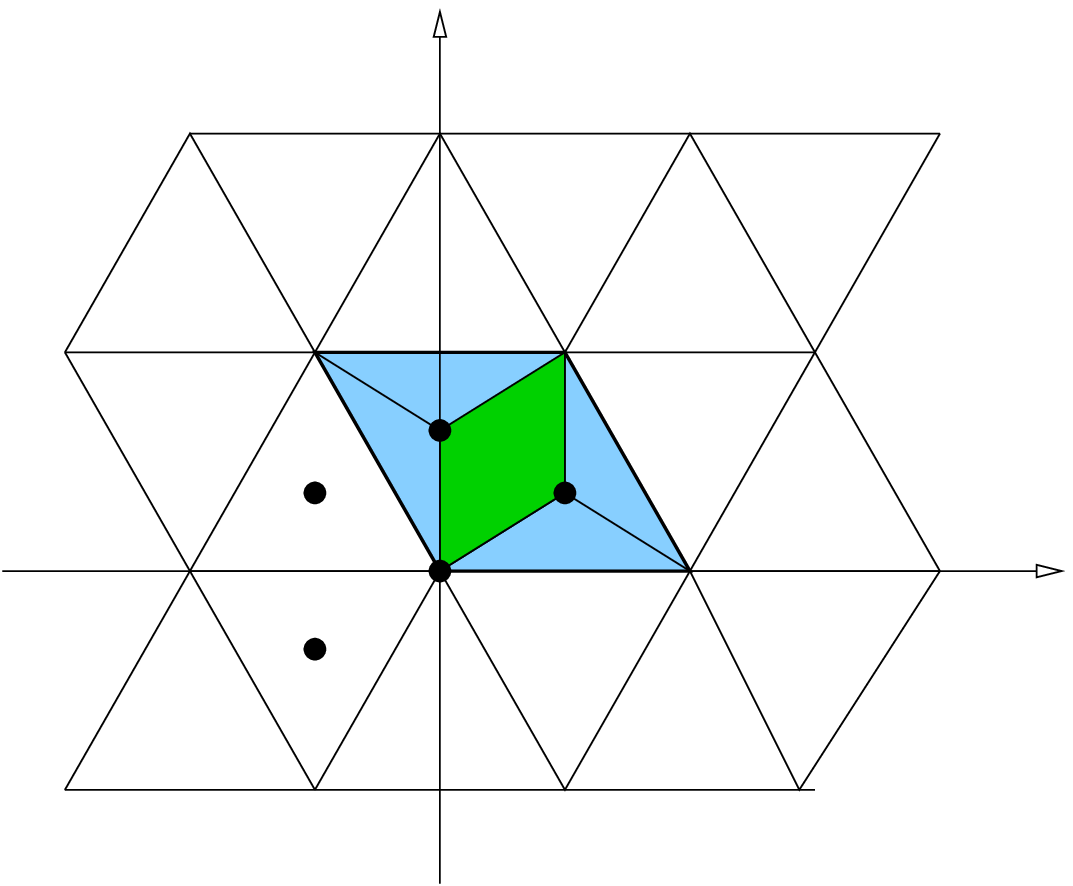,width=.45\linewidth}
\epsfig{file=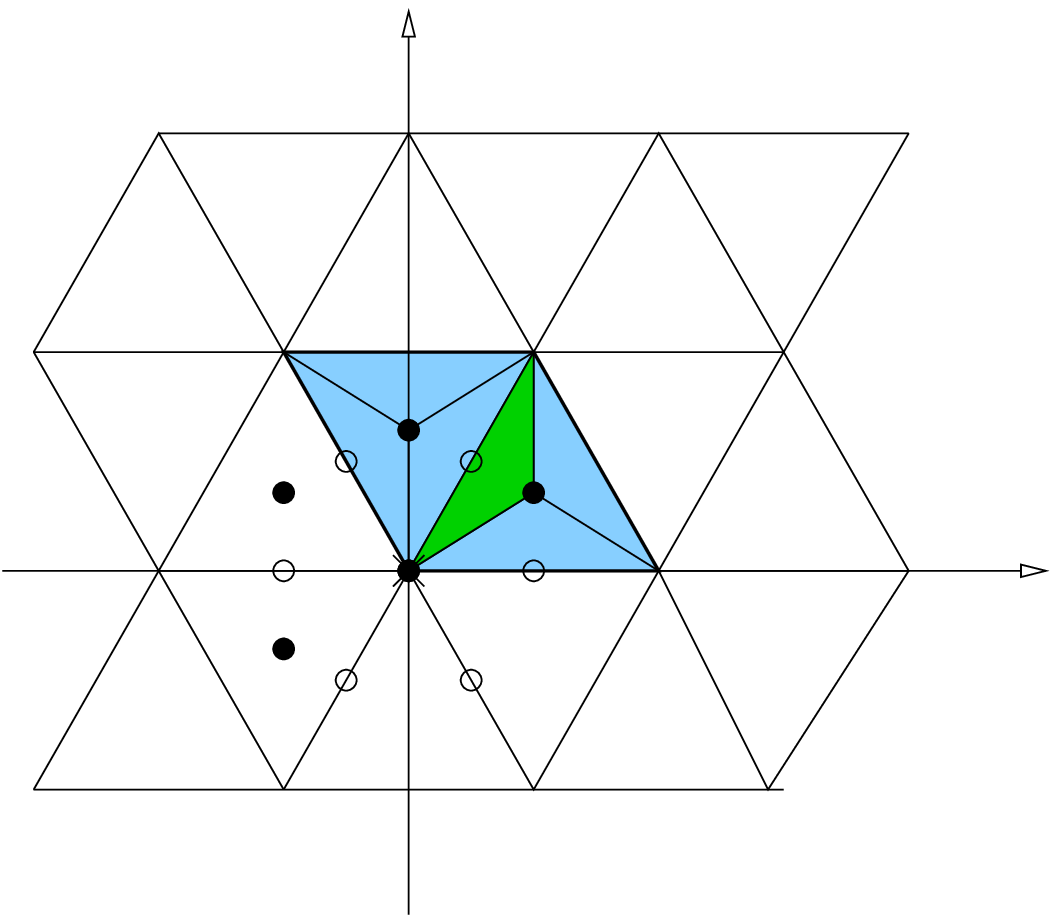,width=.45\linewidth}
\caption{\it $T^2/{\mathbb Z}_3$ (left panel) and $T^2/{\mathbb Z}_6$
(right panel) orbifolds.}
\label{z3z6fig}
\end{figure}

$T^2/{\mathbb Z}_3$ and $T^2/{\mathbb Z}_6$ orbifolds are shown in
Fig.~\ref{z3z6fig}. The fundamental domain of the torus are all shaded
regions while the fundamental domain of the orbifold is only the
darkly shaded (green) one. Open circles correspond to fixed points of
rotations of $\pi$ and filled circels to rotations of $2 \pi/3$. The
only fixed point of the rotations of $\pi/3$ is the origin. The images
of the fixed points are displayed as well and are easyly seen to be
equivalent to its sources by a torus shift.

\end{document}